# The application of adaptive minimum match k-nearest neighbors to identify at-risk students in health professions education

August 9 2022

*Authors*

Anshul Kumar
MGH Institute of Health Professions
(+1) 857.600.1864
[akumar@mghihp.edu](akumar@mghihp.edu)
(first and corresponding author)

Taylor DiJohnson
MGH Institute of Health Professions
(second author)

Roger A. Edwards
MGH Institute of Health Professions
(co-senior author)

Lisa Walker
MGH Institute of Health Professions
(co-senior author)

*Contents*

This document contains the following items, within a single file:
- Article abstract and manuscript (starts on the next page)
- References and other research details
- Supplementary Digital Content: Study appendix

## Abstract

Purpose: When a learner fails to reach a milestone, educators often wonder if there had been any warning signs that could have allowed them to intervene sooner. Machine learning can predict which students are at risk of failing a high-stakes certification exam. If predictions can be made well in advance of the exam, then educators can meaningfully intervene before students take the exam to reduce the chances of a failing score.

Methods: Using already-collected, first-year student assessment data from five cohorts in a Master of Physician Assistant Studies program, the authors implement an "adaptive minimum match" version of the k-nearest neighbors algorithm (AMMKNN), using changing numbers of neighbors to predict each student's future exam scores on the Physician Assistant National Certifying Examination (PANCE). Validation occurred in two ways: Leave-one-out cross-validation (LOOCV) and evaluating the predictions in a new cohort.

Results: AMMKNN achieved an accuracy of 93% in LOOCV. AMMKNN generates a predicted PANCE score for each student, one year before they are scheduled to take the exam. Students can then be classified into extra support, optional extra support, or no extra support groups. The educator then has one year to provide the appropriate customized support to each category of student.

Conclusions: Predictive analytics can identify at-risk students, so they can receive additional support or remediation when preparing for high-stakes certification exams. Educators can use the included methods and code to generate predicted test outcomes for students. The authors recommend that educators use this or similar predictive methods responsibly and transparently, as one of many tools used to support students.

## Introduction

Students in Physician Assistant programs must pass the Physician Assistant National Certifying Examination (PANCE) before they can practice. Obviously, it is the goal of a) students to pass the exam on their first attempt and b) educators to have tools to maximize student success on exams. We use predictive analytics and machine learning methods to develop one such tool and demonstrate how it can be applied within a health professions education program. *The purpose of this tool is to predict (guess) each student's score on a* *high-stakes certification exam one year before the student actually takes the exam.* These predicted scores—if they are accurate enough—can help identify students who might benefit from additional support or remediation from educators before the exam.

We use machine learning to look for patterns in the data of previous cohorts of students, whose certification exam scores are known. We then use these established patterns to make predictions about the future certification exam scores of currently enrolled students. Our data come from five now-graduated cohorts of students in a Master of Physician Assistant (PA) Studies program. Such predictive analytic approaches are used in a variety of educational settings.[1] As Ekowo and Palmer[2] describe, these approaches can be used to both discriminate against or uplift and support vulnerable students, depending on the intentions of the user.

Existing scholarship tends to a) identify at-risk students using standard available machine learning approaches, and b) focus more on model-building than on how to practically leverage predictive modeling when working with students in a health professions program. We build on this work by a) developing and applying a new type of "adaptive minimum match" k-nearest neighbors (AMMKNN) algorithm that uses different numbers of neighbors for each prediction, b) validating our results on an entire cohort of PA students, and c) presenting disaggregated confusion matrices to evaluate predictive results within the context of how they will be applied in practice. Additionally, we have made the AMMKNN algorithm available in an open source R package, called AdaptiveLearnalytics,[3] for others to use and modify.

*Predictive analytics in education*

A typical approach to educational analytics is to use student performance or behavior data at intermediate stages to predict a final outcome of interest. For example, student results on homework assignments or quizzes as well as student activity records in online learning management systems during a semester-long course can be used to make guesses about final exam performance. Prediction of subsequent outcomes based on earlier performance has been demonstrated with students studying a number of topics, including informatics,[4] human computer interaction,[5] computer hardware,[6] and mathematics.[7] Liz-Domınguez et al.[8] refer to this as *grade prediction*. The same approach has also been used to predict which students will withdraw from an educational program.[9,10] These studies train and evaluate the utility of machine learning models. Such machine learning models can be used on future groups of students to create early warning systems that identify at-risk students and intervene, as described by Arnold and Pistilli.[11]

Within health professions specifically, commentaries on machine learning abound,[12–14] while empirical studies that apply machine learning are less common. Black et al.[15] take a similar approach to ours within the physician assistant education context (which we build on by testing additional predictive models, validating our results on an entire new cohort, and addressing practical applications of analytics results). Predictive analytic methods have been used on data from students learning oral pathology,[16] blended medicine,[17] and psychomotor skills.[18] It has also been used to evaluate surgical competence,[19,20] essays written by medical students,[21,22] and residency applicants.[23] Chan and Zary,[24] Tolsgaard et al.,[25] and ten Cate et al.[26] review studies related to analytics in health professions education. All of these examples show that predictive techniques can be used to help educators make guesses about future student outcomes.

## Method

Our study uses data on student performance from a Master of Physician Assistant (PA) Studies program to create a predictive model to identify students at potential risk of failing the Physician Assistant National Certifying Exam (PANCE). Background information about this PA program and data is available in the supplemental *Study Appendix* file (Supplementary Digital Content #1). Table 1 shows the process of data collection, analysis, and application of results as we plan to use it in practice. We used R version 3.6.3 (R Foundation for Statistical Computing, Vienna, Austria) for all analysis. We also developed an open source R package called AdaptiveLearnalytics[3] to carry out key portions of our analysis. Our entire analysis can be found in the *Study Appendix*.

[Table 1]

*Data description and preprocessing*

We utilized de-identified student data from five successive cohorts starting in 2015—a total of 224 students—from a single PA program. The data set was organized such that each row is a student and each column is a variable (containing a measured grade or other characteristic). We utilize iRAT quiz, final exam, final course, and PACKRAT (Physician Assistant Clinical Knowledge Rating and Assessment Tool) scores from students' first year in the PA program, as well as undergraduate grade point average (GPA) data. Even though regular assessments do occur throughout the second year of the PA program, we do not use this data or any demographic data. We exclude second-year data so that an entire year is available to provide additional support to students who are at risk of failing the PANCE.

Our dependent variable is the PANCE score from each student's first attempt at this exam, after the student completes the 25-month PA program. PANCE scores can range from 200–800 points, and students must score 350 or above to pass. We use student performance variables in the data set, described below, as independent variables to predict students' PANCE scores.

This data set, collected as part of this PA program's mostly-TBL curriculum, gives us more information about each student than we would expect to have in a traditional (non-TBL) curriculum. For each TBL course, we calculated the mean of each student's iRAT scores and eliminated the individual iRAT score variables. For example, if students completed seven iRAT quizzes during Course A, we made a new variable with the mean of those seven iRAT quiz scores. We then eliminated the seven individual iRAT quiz scores from our data set.

After these preprocessing steps, we had the following independent variables for each student for 18 first year courses: iRAT mean (when applicable), final exam score, final grade. We also had the following additional variables: overall undergraduate GPA, undergraduate science GPA, first-year PACKRAT score. As described in the *Study Appendix*, we tried two different preparations (A and B) of these independent variables to see which one yielded better results for each type of predictive model.

Once this manual preprocessing of variables was complete, we further narrowed down our independent variables by calculating the Pearson correlation coefficient of each independent variable with PANCE score. Any variable correlated with PANCE below a selected threshold was eliminated. A full explanation and list of variables is in the *Study Appendix*. Prior to training any predictive models, all independent variables were standardized.

*Model selection and adaptive minimum match KNN (AMMKNN)*

Since a numeric score is used to determine if a student passes or fails the PANCE, ours is both a classification and regression problem. It is classification because we are trying to predict if a student passes or fails. It is regression because test results are numeric scores. As educators, we decided it would be most useful to classify students into three groups: very likely to fail (predicted PANCE score <350), moderate risk of failing (predicted PANCE score 350–375), low risk of failing (predicted PANCE score >375). The "moderate risk" category ensures that we were able to identify all "at-risk" students and not only those who would fail. This additional category supports our goal of using predictive analytics to help students advance toward examination with the best preparation resources that we can provide.

Before settling on a single predictive technique, we started by using a number of standard approaches, such as random forest (RF), support vector machine (SVM), and standard K-nearest neighbors (KNN). These attempts are presented in the *Study Appendix*. As shown in our results, none of these allowed us to create a predictive model that would achieve our educational goals of prioritizing the detection of students who might fail the PANCE while still minimizing incorrect (false positive) predictions.

We developed a modified version of KNN called "adaptive minimum match" KNN (AMMKNN), summarized below. If we imagine a student named X who has just finished their first year of the PA program, this is how AMMKNN predicts X's PANCE score:

1. Start with a *training* data set of all previously-graduated students from the program, in which rows are students and columns are the independent variables (the dependent variable, PANCE score, is omitted).
2. Rank the training students from closest to X to farthest from X, based on Euclidean distance.
3. Select the 20 closest matches to X out of the ranked training students. This gives us a list of X's 20 nearest neighbors, to be used in subsequent steps. So far, this procedure is the same as standard KNN with K = 20.
4. We will now use the (known) PANCE scores of these 20 closest matches to make a guess about X's (unknown) future PANCE score.
5. Calculate the mean PANCE score of X's 1 closest neighbor (which is just that neighbor's score itself). Calculate the mean PANCE score of X's 2 closest neighbors. Calculate the mean PANCE score of X's 3 closest neighbors. Continue this process until there are 20 means, each calculated on a different number of X's nearest neighbors. Find the lowest out of the 20 calculated means, called X's *minimum of means*. (Another way to calculate X's minimum of means would be to run the standard KNN procedure 20 times, for all values of K between 1 and 20, yielding 20 different predicted PANCE scores for X; and then choose the lowest predicted score out of the 20 scores).
6. Identify the lowest PANCE score out of X's 20 nearest neighbors. This is X's *minimum match*.

7. Determine if X's PACKRAT (the most important independent variable) score is less or greater than 2 standard deviations below the mean PACKRAT score. (This is to flag students who are possible outliers on the lower end).
    a. If greater (which is common): X's predicted PANCE score is their minimum of means.
    b. If less (which is rare): X's predicted PANCE score is their minimum match.

We can do this procedure each year for Student X and all of their classmates who are in between years 1 and 2 of the PA program. The training data set contains all students from previous cohorts who have completed their first attempt of the PANCE (after they have graduated from the PA program). Standard KNN uses the same value of K—the number of matches—for each student's score prediction. It would then treat the mean of those first K matches as the predicted value for every student. Our adaptive approach differs because it uses a different value of K for each prediction. In the example above, our approach uses a different value of K—based on either the *minimum of means* or *minimum match*—to make a prediction for Student X and each of their classmates. Other adaptive KNN procedures have been used in different contexts.[27–29]

The maximum possible value of K was set at 20 in our study. This means that anywhere between 1 and 20 nearest neighbors could be used to make a prediction. This is further discussed in the *Study Appendix*. The "adaptive minimum match" modification to KNN makes our predicted PANCE scores lower than with standard KNN (and other commonly-used predictive models). This decision was deliberately made to prioritize the detection of students who might fail the PANCE, since standard approaches tend to predict that most students will pass.

*Model evaluation*

We used 181 students' data in our first four cohorts for training and cross validation. We then used our fifth and most recent cohort of 43 students for final validation of our best model (only 42 students are used with variable preparation B, due to missing data for one student). We evaluated the results of all predictive models using leave-one-out cross validation (LOOCV), which has been used or recommended before in similar situations with small sample sizes.[15,30] We adhered as closely as possible to recommendations from Rao et al.[31] For each predictive model, to execute LOOCV on our sample of 181 students, we trained the model on 180 students and tested it on the remaining one student. We repeated this 181 times, such that each student was the testing student one time. This gave us a predicted numeric PANCE score for each student that the computer calculated without "knowing" the true PANCE score of that student. We then classified students as predicted to pass (if their predicted score was 350 or greater) or fail (lower than 350). For each predictive model, we first created a standard 2-by-2 confusion matrix to show each student's predicted value (from when they were the testing student) and their actual PANCE result. From the confusion matrices for each model, we calculated the following standard metrics: true positives, false positives, true negatives, false negatives, accuracy, sensitivity, specificity. Definitions of these metrics are in our *Study Appendix*.

To make our predictions more useful, we further disaggregated them into three groups: predicted to fail (PANCE <350), at-risk of failing (350-375), and likely to pass (>375). This is analogous to the "traffic signal" strategy.[11] We argue that sorting students into these three groups is most useful for our goals of optimizing potential remediation well in advance of the exam. In addition to traditional 2-by-2 confusion matrix data in Table 2, we present 3-by-3 confusion matrices in Figure 1 that show all three groups. We discuss its practical application in our educational context.

## Results

Table 2 shows the LOOCV results of multiple predictive models, based on standard 2-by-2 confusion matrices. AMMKNN has the highest accuracy with both variable preparations (0.93 with preparation B and 0.91 with preparation A). Since our priority as educators is to minimize false negatives (students who "fall through the cracks," meaning they need our help but we fail to detect this), we want to maximize sensitivity while keeping false positives ("unnecessary support" students who are flagged by the model but do not truly need remediation) within reasonable limits. The model with the highest sensitivity is RF(A), which predicts that 35 students will fail, 10 of which are correct predictions and 25 of which are incorrect "unnecessary support" predictions. Providing remediation to 35 students when only 10 of them (less than one-third) need it is not reasonable, we argue. The next highest sensitivity is 0.69, shared by AMMKNN(B) and SVM(B). AMMKNN also has higher accuracy and specificity. AMMKNN(B) predicts that 18 students will fail, with 9 of these predictions being correct and 9 being incorrect. In this scenario, half of all students flagged for extra support would truly require it. SVM(B) predicts that 27 students will fail, with 9 of these predictions being correct and 18 incorrect.

[Table 2]

To more clearly break down the predictions and how we can use them in practice, Figure 1 shows 3-by-3 confusion matrices for selected models. The accuracy for AMMKNN(B) is calculated as the number of correct predictions divided by the number of total students: (9 + 3 + 124)/181 = 0.75. Even though this accuracy is lower than the accuracy of the same model when a 2-by-2 matrix is used (0.93), we argue that the 3-by-3 version—which views results as a spectrum of three classification categories—is more useful when considering the use of the model in practice.

We can build a framework based on the 3-by-3 AMMKNN(B) results: 18 students are predicted to fail the PANCE. Out of these 18 predictions, 9 would be completely correct, 1 would be justifiable (because it would be acceptable to remediate a student who would otherwise almost fail), and 8 would be "unnecessary support" students who receive remediation even though it would not have been needed. The model would identify 26 students as "at risk." Out of these 26 predictions, 5 would be useful (2 students who truly fail and 3 students "at-risk") and 21 would be "unnecessary support." The model would identify 137 students as likely to pass. Out of these 137 predictions, 124 would be "predictable no support" students who pass without risk, 11 would pass with an "at-risk" score, and 2 would "fall through the cracks" because we failed to identify them as needing remediation. The results for AMMKNN(A) are similar, while SVM(B) and standard KNN(B) each have 16 false positives, which is likely too many "unnecessary support" students to reasonably remediate. Our recommended use of these results is to remediate many students with predicted "fail" scores, consider remediating predicted "at-risk" students on a student-by-student basis, and not remediate students with predicted "pass" scores (unless extenuating circumstances or other information suggest otherwise).

[Figure 1]

After completing the validation and inspection of results above, the final step is to further validate our models on an entire cohort of students, the way we intend to do in practice in the future. We do this with our fifth cohort of students, which was not used to train or cross validate the models. Figure 2 shows these results in our 3-by-3 framework. AMMKNN(A) makes the best predictions: Out of 6 students who failed the PANCE, the model correctly predicted that 2 would fail, identified 2 as at-risk, and failed to identify 2 (incorrectly predicting that they would score above 375). Out of 5 students who passed with at-risk scores, the model failed to detect all 5, but this oversight would have been acceptable to us given that the students did not fail the exam. Finally, out of 32 students who passed without risk, 2 were incorrectly predicted to fail, 0 were predicted at-risk, and 30 were correctly predicted to pass.

These results mean that if we had used this model in practice—ignoring other inputs in our discussion below—we would have remediated the 6 students classified as failing or at-risk and we would have been right to remediate 4 out of these 6 students. AMMKNN(B) is the only other model which identifies 4 of the 6 students who fail as either failing or at-risk. SVM and standard KNN fail to detect 4 and 3 students, respectively, out of the 6 total who fail.

The results from both LOOCV and entire-cohort validation suggest that AMMKNN might be more useful in practice than standard SVM and KNN models, because a) AMMKNN performs better in *both* A and B preparations of the independent variables, and b) AMMKNN has fewer false positive predictions at desired levels of sensitivity.

[Figure 2]

## Discussion

Our results demonstrate a) the value of an adaptive machine learning method like AMMKNN that has high sensitivity while minimizing false positives, b) the possible utility of categorizing students into different risk categories, and c) that results from predictive models cannot be trusted blindly without taking into account other relevant factors. We argue that the AMMKNN model can only be useful *in conjunction with* other programmatic information to assist educators in identifying and supporting at-risk students. We argue that the use of predictive models in PA education needs to be considered from three perspectives—educator, student, and program administrator—which together can lead to a practical and ethically-sound student support strategy. Balancing these perspectives involves tradeoffs that remind us that analytics are best used as one among multiple inputs to decision-making. We also acknowledge a number of strengths and limitations of our work in the *Study Appendix*.

*Student considerations*

This study does not include an empirical examination of student reactions to receiving analytics-based results. We do, however, have prior experience in enrolling students in additional support for PANCE preparation, which gives us some of the insights that follow. Prior to incorporating predictive modeling, we used PACKRAT scores to identify at-risk students, a widely used approach.[32–36] Naturally, many of these students were not enthusiastic to find out that they were considered to be at-risk. The terms "at-risk" or "remedial" might sound harsh. We might instead choose to use a term like "needing extra support" or "supplemental curriculum."

Non-academic factors—such as personal, medical, and mental health issues that arise on or before exam day—are important to identify and acknowledge. We fear that students' tendencies to "power through" the scheduled exam rather than rescheduling based on those circumstances could be detrimental in the long run. If a student wakes up on exam day with a migraine and they still decide to take the exam, our machine learning model can never account for this and adjust the prediction for that student. There will always be factors that are not included in the data. We recommend preparing students to be on guard for such circumstances so that they can reschedule their exam. Further research is required on student reactions to (their own) analytics results and non-academic factors in PANCE performance.

*Educator considerations*

Even though educational analytics success stories abound,[2,11] there is also concerning evidence that singling out students as at-risk can be problematic.[37,38] We argue that educators using this and similar predictive models therefore need to be cautious. While providing comprehensive recommendations for this process is beyond the scope of our current work, we note that established scholarship exists related to breaking bad news[39,40] and remediation[41,42] in contexts where students are preparing for high-stakes exams. Framing[43,44] of the results when communicating with the students is something that the educator must consider, so that students understand the motivation and uncertainties associated with suggested actions.

We recommend considering the model's predictions along with other processes established by the program to ensure success of students, including programmatic efforts to support students identified as having unmet needs for a variety of reasons or self-identify as requiring extra support. For example, when we are reviewing the progress of our students each year and deciding which students to recommend for our extra support program, many of the students on this list will have multiple indicators, such as low PACKRAT scores and be identified by the machine learning model. But if the machine learning model identifies additional students who did not have a low PACKRAT score, it at least gives us an opportunity to have a discussion about those students in more depth than we would have otherwise. In practice, taking all information into account, we would then recommend extra support for only a subset of students as they prepare for the PANCE exam.

*Program administrator considerations*

Once a process that incorporates analytics has identified at-risk students, program administrators still face the challenge of identifying and implementing successful interventions. A one-size-fits-all extra-support intervention will not work for everybody, given the unique individual circumstances that students face (discussed above). How do we provide support that meets the needs of the students and stays within the realistic resources of a single program? Administrators must also be prepared for students who are not identified as at-risk raising concerns about equity. In other words, why should some students receive additional support for an incredibly high-stakes exam and not others? Administrators need to balance this equity issue against allocating additional attention selectively.

*Conclusion*

As educators, we hope to create a safety net around our students that can support them as needed, especially in PA education with patient outcomes at stake and delays in certification leading to unwanted professional consequences. The "adaptive minimum match" KNN model we have developed and validated serves as one additional tool that—along with already-existing tools—makes this safety net stronger when used responsibly and transparently. We recommend that analytics should coexist with and bolster other approaches to student support, and that educator, student, and program administrator perspectives should all be incorporated for the practical and ethical implementation of educational analytics.

# Exhibit legends

Table 1: Entire two-year analytics process from program perspective

Table 2: LOOCV predictive model results based on 2-by-2 confusion matrix

Figure 1: LOOCV 3-by-3 confusion matrices for selected predictive models

Figure 2: 3-by-3 confusion matrices for predictions on validation cohort

## Disclosures

*Acknowledgments*

We would like to thank Rupali Khadye-Hadshi and Valay Maskey for their involvement in preparing and formatting this article for publication.

*Funding/Support*

None

*Other disclosures*

None

*Ethical approval*

This research was approved by the Mass General Brigham institutional review board in 2020 (protocol 2020P000514).

*Disclaimers*

None

*Previous presentations*

This article was submitted to the following pre-print servers: https://edarxiv.org/wtuv6/ & https://arxiv.org/abs/2108.07709

## Tables

Table 1: Entire two-year analytics process from program perspective

| **Time →** | Pre-matriculation | Year 1 | End Year 1 | Year 2 | End Year 2 |
|---|---|---|---|---|---|
| **Process step →** | Gather data: Undergraduate GPA | Gather data: iRAT quizzes, final exams, final course grades, PACKRAT score | Run machine learning model to identify at-risk students | Provide additional support to predicted at-risk students. | All students take National Certifying Exam |

Table 2: LOOCV predictive model results based on 2-by-2 confusion matrix

| **Model (variable preparation)** | **TP** | **FP** | **TN** | **FN** | **Accuracy** | **Sensitivity** | **Specificity** |
|---|---|---|---|---|---|---|---|
| Adaptive Minimum Match KNN (B) | 9 | 9 | 159 | 4 | 0.93 | 0.69 | 0.95 |
| Adaptive Minimum Match KNN (A) | 8 | 11 | 157 | 5 | 0.91 | 0.62 | 0.93 |
| Support Vector Machine (B)* | 9 | 18 | 150 | 4 | 0.88 | 0.69 | 0.89 |
| Standard KNN (B)* | 8 | 17 | 151 | 5 | 0.88 | 0.62 | 0.90 |
| Random Forest (A)* | 10 | 25 | 143 | 3 | 0.85 | 0.77 | 0.85 |

Selected best models with each predictive approach are displayed; find all results in *Study Appendix*. Example interpretation: Leave-one-out cross validation (LOOCV) on Adaptive Minimum Match KNN (AMMKNN) with variable preparation A yielded 8 students as true positives (TP), 11 false positives (FP), 157 true negatives (TN), and 5 false negatives (FN).

*These standard models required manual tuning of predicted results (after running the model) to be meaningful, which was not required for AMMKNN. (See *Study Appendix* for additional discussion).

# Figures

Figure 1: LOOCV 3-by-3 confusion matrices for selected predictive models

|  |  | Predicted scores | | |
|---|---|---|---|---|
|  |  | Fail <350 | At-risk 350-375 | Pass >375 |
| Actual scores | <350 | 9 | 2 | 2 |
|  | 350-375 | 1 | 3 | 11 |
|  | >375 | 8 | 21 | 124 |

Figure 1a: Adaptive Minimum Match KNN (B)

|  |  | Predicted scores | | |
|---|---|---|---|---|
|  |  | Fail <350 | At-risk 350-375 | Pass >375 |
| Actual scores | <350 | 9 | 1 | 3 |
|  | 350-375 | 2 | 1 | 12 |
|  | >375 | 16 | 5 | 132 |

Figure 1b: Support Vector Machine (B)

|  |  | Predicted scores | | |
|---|---|---|---|---|
|  |  | Fail <350 | At-risk 350-375 | Pass >375 |
| Actual scores | <350 | 8 | 3 | 2 |
|  | 350-375 | 3 | 4 | 8 |
|  | >375 | 8 | 17 | 128 |

Figure 1c: Adaptive Minimum Match KNN (A)

|  |  | Predicted scores | | |
|---|---|---|---|---|
|  |  | Fail <350 | At-risk 350-375 | Pass >375 |
| Actual scores | <350 | 8 | 2 | 3 |
|  | 350-375 | 1 | 2 | 12 |
|  | >375 | 16 | 20 | 117 |

Figure 1d: Standard KNN (B)

Example interpretation: In the leave-one-out cross validation (LOOCV) results from adaptive minimum match KNN with variable preparation B, 11 students who were predicted to score above 375 actually scored between 350 and 375.

Figure 2: 3-by-3 confusion matrices for predictions on validation cohort

|  |  | Predicted scores | | |
|---|---|---|---|---|
|  |  | Fail <350 | At-risk 350-375 | Pass >375 |
| Actual scores | <350 | 2 | 2 | 2 |
|  | 350-375 | 0 | 2 | 3 |
|  | >375 | 3 | 2 | 26 |

Figure 2a: Adaptive Minimum Match KNN (B)

|  |  | Predicted scores | | |
|---|---|---|---|---|
|  |  | Fail <350 | At-risk 350-375 | Pass >375 |
| Actual scores | <350 | 2 | 0 | 4 |
|  | 350-375 | 2 | 0 | 3 |
|  | >375 | 0 | 0 | 31 |

Figure 2b: Support Vector Machine (B)

|  |  | Predicted scores | | |
|---|---|---|---|---|
|  |  | Fail <350 | At-risk 350-375 | Pass >375 |
| Actual scores | <350 | 2 | 2 | 2 |
|  | 350-375 | 0 | 0 | 5 |
|  | >375 | 2 | 0 | 30 |

Figure 2c: Adaptive Minimum Match KNN (A)

|  |  | Predicted scores | | |
|---|---|---|---|---|
|  |  | Fail <350 | At-risk 350-375 | Pass >375 |
| Actual scores | <350 | 2 | 1 | 3 |
|  | 350-375 | 1 | 1 | 3 |
|  | >375 | 0 | 2 | 29 |

Figure 2d: Standard KNN (B)

Example interpretation: In the full-cohort validation results from adaptive minimum match KNN with variable preparation B, there were 0 students who were predicted to score below 350 and actually scored between 350 and 375.

# Study Appendix: The application of adaptive minimum match k-nearest neighbors to identify at-risk students in health professions education

Anshul Kumar, Taylor DiJohnson, Lisa Walker, Roger Edwards

Version – 05 August 2022

## Contents

# 1   Information and initial set-up

This supplementary *Study Appendix* file accompanies the article *The application of adaptive minimum match k-nearest neighbors to identify at-risk students in health professions education* in 2022. This file was produced using R Markdown in RStudio.

This file shows the creation of the best RF, SVM, KNN, and adaptive minimum match KNN models that we were able to create. Other attempts for each type of model are not shown.

**Corresponding author: Anshul Kumar, akumar@mghihp.edu**

## 1.1 Background on physician assistant education and team-based learning

The physician assistant (PA) profession is currently undergoing tremendous growth in the United States. To become a certified PA in the United States, students typically study for two or more years and then must achieve a score of at least 350 on the Physician Assistant National Certifying Exam (PANCE; graded on a 200–800 scale). In a typical PA program, the first year of education primarily features coursework with regular assessments and the second year consists of clinical rotations. Given these characteristics, PA programs are a good example of a health professions program that might benefit from predictive analytic tools.

Our data come from a PA program that mostly uses team-based learning (TBL), which is relatively new in PA education and has been growing in popularity in medical and other health professions programs.[1] TBL appears to be used at some medical schools at pertinent stages of the curriculum.[2] [3]

TBL relies on small group interaction, creating opportunities for students to solve problems together and practice key concepts.[4] The learning process consists of three stages: student preparation, readiness assurance, and application.[5] Wallace and Walker[6] describe the following key details of TBL: a) Create TBL assessments and learning activities using backwards design, b) Form diverse teams of 5–7 students, c) Students study content before class. In class, they engage in individual and team assessments, listen to minilectures, and participate in application activities. TBL assessments include iRATs (individual readiness assurance test) and tRATs (team readiness assurance test). Students in this PA program also take the PACKRAT (Physician Assistant Clinical Knowledge Rating and Assessment Tool) exam at the end of their first year in the program. The PACKRAT is a cumulative, nationally-benchmarked assessment that covers important PA knowledge areas.

The frequency of assessments in TBL curricula generates data that can be easily included in predictive models. However, such an abundance of data is by no means necessary for using our methods. In fact, we ended up combining or eliminating many pieces of information. Health professions education programs that do not generate as much data can still use predictive analytics.

## 1.2 Initial set-up

The R code for many of the functions/processes used in this study are stored in the R package that we have created for this analysis, called `AdaptiveLearnalytics`. Below, we install and load this package.

```
if (!require(devtools)) install.packages('devtools')
library(devtools)
devtools::install_github("readcreate/AdaptiveLearnalytics")
```

```r
library(AdaptiveLearnalytics)
```

```
##
## Thanks for using the AdaptiveLearnalytics package.
##
## Please cite this package when you use it.
##
## We welcome your questions or feedback: Anshul Kumar <akumar@mghihp.edu>
##
## Suggested citation:
## Anshul Kumar and Roger Edwards. 2021. AdaptiveLearnalytics: Adaptive Predictive Learning Analytic Too
##
## Run the following code for more information:
## help(package = 'AdaptiveLearnalytics')
##
## This package was initially developed to assist with the following publication: Anshul Kumar, Lisa Wal
##
## We used the following resources to make this package:
## * Erik Erhardt. 2018. R Package Development. https://statacumen.com/teach/ShortCourses/R_Packages/R_
## * Hadley Wickham. R Packages. https://r-pkgs.org/
## * Sharon Machlis. 2019. How to write your own R package. IDG TECHtalk. https://www.youtube.com/watch
```

Readers can use the code above to install and use the package.

The following code can be run to see more information:

```r
help(package = 'AdaptiveLearnalytics')
```

- Below, we set our working directory:

```r
setwd("path/to/working/directory")
```

The code above is for illustrative purposes. We set our real working directory in a hidden code chunk.

We are unfortunately not able to make our data publicly available. However, the code below combined with the examples within the `AdaptiveLearnalytics` package should be sufficient to reproduce our analysis methods on a different data set.

# 2 Prepare and explore data

## 2.1 Load data from Excel file

The code below subsets our data and produces output to help us confirm that the subsetting worked.

```r
yearcutoff <- 2019 # writing 2018 here will include first 3 cohorts only

# load data into R
library(readxl)
d <- read_excel("AI Research Template downloaded 20220207.xlsx")

dpartial <- d[c("Year student began program","PANCE")]

dim(d)
```

```r
## [1] 323 426

d <- d[which(d$`Year student began program`<yearcutoff),]

table(d$`Year student began program`)
```

```
## 
## 2015 2016 2017 2018 
##   40   49   50   44
```

```r
dim(d)
```

```
## [1] 183 426
```

```r
d$missingPANCE <- ifelse(is.na(d$PANCE), "missing","present")

d <- d[which(!is.na(d$PANCE)),]

dim(d)
```

```
## [1] 181 427
```

```r
masteralumnidata <- d
```

## 2.2 Explore dependent variable (PANCE) results

The dependent variable in this study is student results on the Physician Assistant National Certifying Examination (PANCE). The independent variables are the results of quizzes, tests, and other assessments that physician assistant students engage in during their studies to become a physician assistant.

PANCE results by year:

```r
dpartial$pass <- ifelse(dpartial$PANCE>349,1,0)
dpartial$fail <- ifelse(dpartial$PANCE>349,0,1)

library(dplyr)

dplyr::group_by(dpartial, `Year student began program`) %>%
dplyr::summarise(
  count = n(),
  mean = mean(PANCE, na.rm = TRUE),
  sd = sd(PANCE, na.rm = TRUE),
  pass = sum(pass, na.rm = TRUE),
  fail = sum(fail, na.rm = TRUE)
  )
```

```
## # A tibble: 7 x 6
##   `Year student began program` count  mean    sd  pass  fail
##                          <dbl> <int> <dbl> <dbl> <dbl> <dbl>
## 1                         2015    40  417.  65.6    35     5
## 2                         2016    49  440.  51.4    49     0
```

```
## 3                           2017  50 431.  73.9      43       7
## 4                           2018  44 445.  61.9      41       1
## 5                           2019  46 434.  75.6      37       6
## 6                           2020  46 NaN   NA         0       0
## 7                           2021  48 NaN   NA         0       0
```

Example interpretation: In 2015, we had a cohort of 40 students. Their mean score on the PANCE was 416.7 points with a standard deviation of 65.6 points. On their first attempt, 35 students passed the exam and 5 failed.

Note that in some cohorts, not all students went on to take the PANCE.

In our study, we use the four years of student data from 2015 through 2018 to build and cross-validate our models. We refer to students in the 2015–2018 cohorts as "alumni."

We then use the 2019 cohort—who we often refer to as the "current students," because this model is eventually meant to be applied to a cohort of students who have not yet graduated—for further validation. 2019 students are never used to train a predictive model; they are only used for validation of the various models we create, to simulate the way that we hope to apply our best model(s) to an entire cohort at once.

## 2.3 Prepare alumni data (for LOOCV)

### 2.3.1 Details

There are two possible ways to prepare the data, which we will call preparations A and B (these might also be referred to as `.a` and `.b` and be used as suffixes within the code). The `.a` preparation does not use the CAT variables (the final exams for the courses) overall, with one exception, as shown below. The `.a` preparation then uses a lower correlation threshold for variable selection than the `.b` preparation.

### 2.3.2 Preparation A – alumni

```
d <- masteralumnidata
```

Courses in final semester before rotations:

- PA722: PA in the community
- PA738: Essentials of gastroenterology
- PA739: Essentials of endocrinology
- PA740: Essentials of nephrology and urology
- PA741: Principles of reproductive medicine
- PA752: Patient Care III
- PA760: Special populations
- PA770: Principles of surgery, emergency, and inpatient care

```
sd(d$`PA722  PA in  Comm Final Grade`)
```

```
## [1] 0.04587185
```

```
sd(d$`PA738 Gastroenterlogy Final Grade`)
```

```
## [1] 0.04442174
```

```r
sd(d$`PA739  Endocrinology Final Grade`)
```

```
## [1] 0.04176545
```

```r
sd(d$`PA740 Neph& Uro Final`)
```

```
## [1] 0.04307088
```

```r
sd(d$`PA741 Prin  of Repro Med Final Grade`)
```

```
## [1] 0.03776317
```

```r
sd(d$`PA752 Patient Care III Final Grade`)
```

```
## [1] 0.02231169
```

```r
sd(d$`PA760 Special Pops Final Grade`)
```

```
## [1] 0.03017262
```

```r
sd(d$`PA770 Surg, EM, and IP Care Final Grade`)
```

```
## [1] 0.0363315
```

```r
summary(d$`PA722  PA in  Comm Final Grade`)
```

```
##    Min. 1st Qu.  Median    Mean 3rd Qu.    Max. 
##  0.7579  0.9684  1.0000  0.9730  1.0000  1.0000
```

```r
summary(d$`PA738 Gastroenterlogy Final Grade`)
```

```
##    Min. 1st Qu.  Median    Mean 3rd Qu.    Max. 
##  0.7466  0.8789  0.9080  0.9015  0.9320  0.9800
```

```r
summary(d$`PA739  Endocrinology Final Grade`)
```

```
##    Min. 1st Qu.  Median    Mean 3rd Qu.    Max. 
##  0.7770  0.8703  0.9004  0.8977  0.9278  0.9863
```

```r
summary(d$`PA740 Neph& Uro Final`)
```

```
##    Min. 1st Qu.  Median    Mean 3rd Qu.    Max. 
##  0.7792  0.8730  0.9008  0.9005  0.9273  0.9867
```

```r
summary(d$`PA741 Prin  of Repro Med Final Grade`)
```

```
##    Min. 1st Qu.  Median    Mean 3rd Qu.    Max.
##  0.7727  0.8715  0.8970  0.8962  0.9226  0.9810
```

```r
summary(d$`PA752 Patient Care III Final Grade`)
```

```
##    Min. 1st Qu.  Median    Mean 3rd Qu.    Max.
##  0.8860  0.9446  0.9590  0.9559  0.9740  0.9930
```

```r
summary(d$`PA760 Special Pops Final Grade`)
```

```
##    Min. 1st Qu.  Median    Mean 3rd Qu.    Max.
##  0.8002  0.9390  0.9568  0.9526  0.9733  0.9980
```

```r
summary(d$`PA770 Surg, EM, and IP Care Final Grade`)
```

```
##    Min. 1st Qu.  Median    Mean 3rd Qu.    Max.
##  0.7534  0.8532  0.8763  0.8744  0.8964  0.9536
```

Based on the summary statistics above, we see that PA770 is likely the hardest final semester class (large range and standard devition), so we include the CAT (final exam) for this class as an independent variable in our model, as an additional indicator from each student's final semester of courses.

Below, we remove a few variables that we know we definitely don't want to include (due to data being unavailable, specific variable coding, or our own hypotheses and experience about which variables will lead to the best predictions).

```r
# Variables were manually selected

d <- d[c(
  # "cohort",
  "Overall GPA"
  ,"Overall Science GPA"
  # ,"GRE Official Overall Score"
  # ,"GRE Analytical Scaled"
  # ,"GRE Quantitative Scaled"
  # ,"GRE Verbal Scaled"
  ,"Foundations IRAT Mean"
  # ,"Foundations CAT"
  ,"PA650 Foundations  of Med Final Grade" # Count courses. Course 1
  # ,"PA Professions Exam 1"
  # ,"PA Professions Exam 2"
  # ,"Hematology iRAT Mean" # removed from curriculum
  # ,"Hematology Online Anatomy Practical" # removed from curriculum
  # ,"Hematology CAT" # removed from curriculum
  # ,"PA732 Hematology Final Grade" # removed from curriculum
  ,"Derm iRAT Mean"
  # ,"Derm CAT"
  #, derm practical only started 2019 so can't include
  ,"PA730 Derm Final Grade" # Course 2
```

```
,"MSK iRAT Mean"
# ,"Practical 1"
# practical 2 no longer happens
# ,"MSK CAT"
,"PA731 Muscskl Ds & Injury Final Grade" # Course 3
,"Neuro iRAT Mean"
# ,"Neuro Anatomy Practical"
# ,"Neuro CAT"
,"PA734  Neurology Final Grade" # Course 4
# ,"Patient Care Final Written Exam (CAT 1)"
# leaving out PCI practicals almost all students got 100%
,"Patient Care I Final Grade"
# ,"PA in Practice Practical Exam 1" # need to add
# ,"PA in Practice coding and billing quiz" # problematic
# ,"PA in Practice Written Exam 1" # similar to CAT
# ,"PA in Practice  Practical Exam 2" # need to add
# ,"PA in Practice Written Exam 2" # similar to CAT
# ,"PA in Practice Exam 4 Practical" # not a thing
,"PA721  PA in Practice  Final Grade" # Course 5
,"Cardio iRAT Mean"
# ,"Cardio CAT #1"
# ,"Cardio Anatomy Practical" # need to add
,"PA736 Cardiovascular Disease Final Grade" # Course 6
,"Pulm iRAT Mean"
# ,"Pulm Anatomy Practical" # class of 2019 didn't take it
# ,"Pulm CAT"
,"PA733 Pulmonary Med Final Grade" # Course 7
,"Otolar & Ophthal iRAT Mean"
# ,"Otolar & OphthalAnatomy Practical"
# ,"Otolar & OphthalCAT"
,"PA737 Otolar & Ophthal Final Grade" # Course 8
,"Behav Med iRAT Mean"
# ,"Behav Med CAT"
,"PA735 Prin of Beh  Med Final Grade" # Course 9
# ,"Patient Care II Exam 1"
#,  PCII practical omitted all students 100%
,"PA751 Patient Care II Final Grade" # Course 10
,"PA722  PA in  Comm Final Grade" # Course 11
,"Neph/Ur iRAT Mean"
# ,"Neph/Uro CAT"
# ,"Neph/Uro Anatomy Practical" # not everyone took it
,"PA740 Neph& Uro Final" # Course 12
# ,"Endo Anatomy Practical" # need to add
# ,"Endo CAT"
,"PA739  Endocrinology Final Grade" # Course 13
,"GI iRAT Mean"
# ,"GI CAT"
# ,"GI Anatomy Practical"
,"PA738 Gastroenterlogy Final Grade" # Course 14
,"Repro iRAT Mean"
# ,"Repro Anatomy Practical"
# ,"Repro CAT"
,"PA741 Prin  of Repro Med Final Grade" # Course 15
```

```r
   # ,"Pt Care III Exam 1" # maybe eliminated
   #, omitting PCIII practicals since almost all 100%
   # , "Pt Care III Practical Exam 1" # not everyone took it
   # , "Pt Care III Practical Exam 2" # not everyone took it
   ,"PA752 Patient Care III Final Grade" # Course 16
   ,"Special Pops Midterm" # no iRATs so included
   ,"Special Pops Final"
   ,"PA760 Special Pops Final Grade" # Course 17
   ,"SEI iRAT Mean"
   ,"SEI CAT" # included because hardest exam in final semester
   ,"PA770 Surg, EM, and IP Care Final Grade" # Course 18
   ,"PACKRAT I  Raw Score"
   # ,"PACKRAT I Cardio"
   # ,"PACKRAT I Derm"
   # ,"PACKRAT I Endo"
   # ,"PACKRAT I ENT/Opht"
   # ,"PACKRAT I GI"
   # ,"PACKRAT I HEMATOLOGY"
   # ,"PACKRAT I INFECTIOUS DISEASES"
   # ,"PACKRAT I NEURO"
   # ,"PACKRAT I OBGYN"
   # ,"PACKRAT I ORTHO/RHEM"
   # ,"PACKRAT I PSYCH/BEHVMED"
   # ,"PACKRAT I PULM"
   # ,"PACKRAT I URO/REAL"
   # ,"PACKRAT I CLINICAL INTERVENTION"
   # ,"PACKRAT I CLINICAL THERAPEUTICS"
   # ,"PACKRAT I DIAGNOSIS"
   # ,"PACKRAT I DIAGNOSTIC STUDIES"
   # ,"PACKRAT I HEALTH MAINTENANCE"
   # ,"PACKRAT 1 HISTORY & PHYSICAL"
   # ,"PACKRAT I SCIENTIFIC CONCEPTS"
   # ,"PA820-828 Family Medicine EOR" # most related to PANCE
   # ,"PA820-828 Internal Medicine EOR" # most related to PANCE
   # ,"PA820-828 Pediatrics EOR"
   # ,"PA820-828 Women's Health EOR"
   # ,"PA820-828 Psychiatry EOR"
   # ,"PA820-828 Emergency Med EOR"
   # ,"PA820-828 General Surgery EOR"
   # ,"PACKRAT II Raw Score"
   # ,"Remediated"
   ,"PANCE"
)]

names(d) <- make.names(names(d))

library(jtools)
df<-jtools::standardize(d)
dfcopystd <- df

# put unstandardized dependent variable back into data
df$PANCE <- d$PANCE
```

```r
CorrelationThreshold <- .1
```

We want to remove variables from the data set that are correlated at lower than 0.1 with the dependent variable `PANCE`. Following standard practice in machine learning approaches, we tried many different configurations of independent variables and many different correlation thresholds and found 0.1 to yield the best list of variables to make predictions. We decided what list of variables was "best" by comparing predictive model results with different lists of variables, using accuracy, sensitivity, and specificity to decide.

We are not able to show every single attempt in this file, but the two variable configurations that we present in this file demonstrate our general approach: try different sets of variables in all of the machine learning models, then choose only the best performing models (regardless of the variables) to validate with our 2019 cohort of "current students" and use in our own educational work on new students. This method of comparing and selecting models appears to be a standard procedure, also practiced by many of the scholars we cite in our main article's literature review.

```r
df <- corVarSelect("PANCE",df,CorrelationThreshold)
```

```
## 
## 
## Starting process. Input dataframe is called df and contains 181 observations and 38 variables, inclu
## 
## 
## 
## 1. Correlation between Overall.GPA (SD = 1) and PANCE = 0.2387433.
## 
## 2. Correlation between Overall.Science.GPA (SD = 1) and PANCE = 0.2679229.
## 
## 3. Correlation between Foundations.IRAT.Mean (SD = 1) and PANCE = 0.4735913.
## 
## 4. Correlation between PA650.Foundations..of.Med.Final.Grade (SD = 1) and PANCE = 0.40957.
## 
## 5. Correlation between Derm.iRAT.Mean (SD = 1) and PANCE = 0.3362847.
## 
## 6. Correlation between PA730.Derm.Final.Grade (SD = 1) and PANCE = 0.3167553.
## 
## 7. Correlation between MSK.iRAT.Mean (SD = 1) and PANCE = 0.349823.
## 
## 8. Correlation between PA731.Muscskl.Ds...Injury.Final.Grade (SD = 1) and PANCE = 0.3713423.
## 
## 9. Correlation between Neuro.iRAT.Mean (SD = 1) and PANCE = 0.4803686.
## 
## 10. Correlation between PA734..Neurology.Final.Grade (SD = 1) and PANCE = 0.4156433.
## 
## 11. Correlation between Patient.Care.I.Final.Grade (SD = 1) and PANCE = 0.2582716.
## 
## 12. Correlation between PA721..PA.in.Practice..Final.Grade (SD = 1) and PANCE = 0.03645805.
## 
## 13. Correlation between Cardio.iRAT.Mean (SD = 1) and PANCE = 0.5121714.
## 
## 14. Correlation between PA736.Cardiovascular.Disease.Final.Grade (SD = 1) and PANCE = 0.5350263.
## 
## 15. Correlation between Pulm.iRAT.Mean (SD = 1) and PANCE = 0.4573783.
## 
## 16. Correlation between PA733.Pulmonary.Med.Final.Grade (SD = 1) and PANCE = 0.5404144.
```

```
## 
## 17. Correlation between Otolar...Ophthal.iRAT.Mean (SD = 1) and PANCE = 0.4394858.
## 
## 18. Correlation between PA737.Otolar...Ophthal.Final.Grade (SD = 1) and PANCE = 0.3156686.
## 
## 19. Correlation between Behav.Med.iRAT.Mean (SD = 1) and PANCE = 0.3885394.
## 
## 20. Correlation between PA735.Prin.of.Beh..Med.Final.Grade (SD = 1) and PANCE = 0.3518571.
## 
## 21. Correlation between PA751.Patient.Care.II.Final.Grade (SD = 1) and PANCE = 0.2672445.
## 
## 22. Correlation between PA722..PA.in..Comm.Final.Grade (SD = 1) and PANCE = -0.05965598.
## 
## 23. Correlation between Neph.Ur.iRAT.Mean (SD = 1) and PANCE = 0.2687676.
## 
## 24. Correlation between PA740.Neph..Uro.Final (SD = 1) and PANCE = 0.3217865.
## 
## 25. Correlation between PA739..Endocrinology.Final.Grade (SD = 1) and PANCE = 0.4201746.
## 
## 26. Correlation between GI.iRAT.Mean (SD = 1) and PANCE = 0.2251731.
## 
## 27. Correlation between PA738.Gastroenterlogy.Final.Grade (SD = 1) and PANCE = 0.4150282.
## 
## 28. Correlation between Repro.iRAT.Mean (SD = 1) and PANCE = 0.431119.
## 
## 29. Correlation between PA741.Prin..of.Repro.Med.Final.Grade (SD = 1) and PANCE = 0.3940514.
## 
## 30. Correlation between PA752.Patient.Care.III.Final.Grade (SD = 1) and PANCE = 0.2629877.
## 
## 31. Correlation between Special.Pops.Midterm (SD = 1) and PANCE = 0.2966964.
## 
## 32. Correlation between Special.Pops.Final (SD = 1) and PANCE = 0.278615.
## 
## 33. Correlation between PA760.Special.Pops.Final.Grade (SD = 1) and PANCE = 0.3704975.
## 
## 34. Correlation between SEI.iRAT.Mean (SD = 1) and PANCE = 0.4279774.
## 
## 35. Correlation between SEI.CAT (SD = 1) and PANCE = 0.5089996.
## 
## 36. Correlation between PA770.Surg..EM..and.IP.Care.Final.Grade (SD = 1) and PANCE = 0.5675375.
## 
## 37. Correlation between PACKRAT.I..Raw.Score (SD = 1) and PANCE = 0.6443328.
## 
## 38. Correlation between PANCE (SD = 64.05958) and PANCE = 1.
## 
## Variable selection process is complete. Returned dataframe contains 181 observations and 36 variable
```

Make a copy of the new data set, with correlation-selected variables:

```
FinalAlumni.a <- df
```

Final list of variables (also save a copy)

```r
names(FinalAlumni.a)
```

```
##  [1] "Overall.GPA"
##  [2] "Overall.Science.GPA"
##  [3] "Foundations.IRAT.Mean"
##  [4] "PA650.Foundations..of.Med.Final.Grade"
##  [5] "Derm.iRAT.Mean"
##  [6] "PA730.Derm.Final.Grade"
##  [7] "MSK.iRAT.Mean"
##  [8] "PA731.Muscskl.Ds...Injury.Final.Grade"
##  [9] "Neuro.iRAT.Mean"
## [10] "PA734..Neurology.Final.Grade"
## [11] "Patient.Care.I.Final.Grade"
## [12] "Cardio.iRAT.Mean"
## [13] "PA736.Cardiovascular.Disease.Final.Grade"
## [14] "Pulm.iRAT.Mean"
## [15] "PA733.Pulmonary.Med.Final.Grade"
## [16] "Otolar...Ophthal.iRAT.Mean"
## [17] "PA737.Otolar...Ophthal.Final.Grade"
## [18] "Behav.Med.iRAT.Mean"
## [19] "PA735.Prin.of.Beh..Med.Final.Grade"
## [20] "PA751.Patient.Care.II.Final.Grade"
## [21] "Neph.Ur.iRAT.Mean"
## [22] "PA740.Neph..Uro.Final"
## [23] "PA739..Endocrinology.Final.Grade"
## [24] "GI.iRAT.Mean"
## [25] "PA738.Gastroenterlogy.Final.Grade"
## [26] "Repro.iRAT.Mean"
## [27] "PA741.Prin..of.Repro.Med.Final.Grade"
## [28] "PA752.Patient.Care.III.Final.Grade"
## [29] "Special.Pops.Midterm"
## [30] "Special.Pops.Final"
## [31] "PA760.Special.Pops.Final.Grade"
## [32] "SEI.iRAT.Mean"
## [33] "SEI.CAT"
## [34] "PA770.Surg..EM..and.IP.Care.Final.Grade"
## [35] "PACKRAT.I..Raw.Score"
## [36] "PANCE"
```

```r
CorSelectedVars.a <- names(FinalAlumni.a)
saveRDS(CorSelectedVars.a, file = "CorSelectedVars.a.rds")
```

### 2.3.3 Preparation B – alumni

This version of the data preparation, called `.b`, leaves in the CAT exams (the final exams in the various courses) and has a higher correlation threshold for variable selection than the `.a` preparation.

```r
d<-masteralumnidata

d <- d[c(
  # "cohort",
  "Overall GPA"
```

```
,"Overall Science GPA"
# ,"GRE Official Overall Score"
# ,"GRE Analytical Scaled"
# ,"GRE Quantitative Scaled"
# ,"GRE Verbal Scaled"
,"Foundations IRAT Mean"
,"Foundations CAT"
,"PA650 Foundations  of Med Final Grade"
,"PA Professions Exam 1"
,"PA Professions Exam 2"
# ,"Hematology iRAT Mean" # MUST remove since no longer in curriculum
# ,"Hematology Online Anatomy Practical"
# ,"Hematology CAT"
# ,"PA732 Hematology Final Grade"
,"Derm iRAT Mean"
,"Derm CAT"
,"PA730 Derm Final Grade"
,"MSK iRAT Mean"
,"Practical 1"
,"MSK CAT"
,"PA731 Muscskl Ds & Injury Final Grade"
,"Neuro iRAT Mean"
,"Neuro Anatomy Practical"
,"Neuro CAT"
,"PA734  Neurology Final Grade"
,"Patient Care Final Written Exam (CAT 1)"
,"Patient Care I Final Grade"
# ,"PA in Practice coding and billing quiz" # problematic
,"PA in Practice Written Exam 1"
,"PA in Practice  Practical Exam 2"
,"PA in Practice Written Exam 2"
# ,"PA in Practice Exam 4 Practical" # problematic
,"PA721  PA in Practice  Final Grade"
,"Cardio iRAT Mean"
,"Cardio CAT #1"
,"PA736 Cardiovascular Disease Final Grade"
,"Pulm iRAT Mean"
,"Pulm CAT"
,"PA733 Pulmonary Med Final Grade"
,"Otolar & Ophthal iRAT Mean"
,"Otolar & OphthalAnatomy Practical"
,"Otolar & OphthalCAT"
,"PA737 Otolar & Ophthal Final Grade"
,"Behav Med iRAT Mean"
,"Behav Med CAT"
,"PA735 Prin of Beh  Med Final Grade"
,"Patient Care II Exam 1"
,"PA751 Patient Care II Final Grade"
,"Neph/Ur iRAT Mean"
,"Neph/Uro CAT"
,"Endo Anatomy Practical"
,"Endo CAT"
,"PA739  Endocrinology Final Grade"
```

```r
  ,"GI iRAT Mean"
  ,"GI CAT"
  ,"GI Anatomy Practical"
  ,"PA738 Gastroenterlogy Final Grade"
  ,"Repro iRAT Mean"
  ,"Repro Anatomy Practical"
  ,"Repro CAT"
  ,"PA741 Prin  of Repro Med Final Grade"
  ,"Pt Care III Exam 1"
  ,"PA752 Patient Care III Final Grade"
  ,"Special Pops Midterm"
  ,"Special Pops Final"
  ,"PA760 Special Pops Final Grade"
  ,"SEI iRAT Mean"
  ,"SEI CAT" # formerly "SEI Written Exam"
  ,"PA770 Surg, EM, and IP Care Final Grade"
  ,"PACKRAT I  Raw Score"
  # ,"PACKRAT I Cardio"
  # ,"PACKRAT I Derm"
  # ,"PACKRAT I Endo"
  # ,"PACKRAT I ENT/Opht"
  # ,"PACKRAT I GI"
  # ,"PACKRAT I HEMATOLOGY"
  # ,"PACKRAT I INFECTIOUS DISEASES"
  # ,"PACKRAT I NEURO"
  # ,"PACKRAT I OBGYN"
  # ,"PACKRAT I ORTHO/RHEM"
  # ,"PACKRAT I PSYCH/BEHVMED"
  # ,"PACKRAT I PULM"
  # ,"PACKRAT I URO/REAL"
  # ,"PACKRAT I CLINICAL INTERVENTION"
  # ,"PACKRAT I CLINICAL THERAPEUTICS"
  # ,"PACKRAT I DIAGNOSIS"
  # ,"PACKRAT I DIAGNOSTIC STUDIES"
  # ,"PACKRAT I HEALTH MAINTENANCE"
  # ,"PACKRAT 1 HISTORY & PHYSICAL"
  # ,"PACKRAT I SCIENTIFIC CONCEPTS"
  ,"PANCE"
)]

names(d) <- make.names(names(d))

library(jtools)
df<-jtools::standardize(d)
dfcopystd <- df

# put unstandardized dependent variable back into data
df$PANCE <- masteralumnidata$PANCE

CorrelationThreshold <- .19
```

We want to remove variables from the data set that are correlated at lower than 0.19 with the dependent variable `PANCE`.

```r
df <- corVarSelect("PANCE",df,CorrelationThreshold)
```

```
## 
## 
## Starting process. Input dataframe is called df and contains 181 observations and 62 variables, inclu
## 
## 
## 
## 1. Correlation between Overall.GPA (SD = 1) and PANCE = 0.2387433.
## 
## 2. Correlation between Overall.Science.GPA (SD = 1) and PANCE = 0.2679229.
## 
## 3. Correlation between Foundations.IRAT.Mean (SD = 1) and PANCE = 0.4735913.
## 
## 4. Correlation between Foundations.CAT (SD = 1) and PANCE = 0.3551673.
## 
## 5. Correlation between PA650.Foundations..of.Med.Final.Grade (SD = 1) and PANCE = 0.40957.
## 
## 6. Correlation between PA.Professions.Exam.1 (SD = 1) and PANCE = 0.117246.
## 
## 7. Correlation between PA.Professions.Exam.2 (SD = 1) and PANCE = -0.02488566.
## 
## 8. Correlation between Derm.iRAT.Mean (SD = 1) and PANCE = 0.3362847.
## 
## 9. Correlation between Derm.CAT (SD = 1) and PANCE = 0.3340736.
## 
## 10. Correlation between PA730.Derm.Final.Grade (SD = 1) and PANCE = 0.3167553.
## 
## 11. Correlation between MSK.iRAT.Mean (SD = 1) and PANCE = 0.349823.
## 
## 12. Correlation between Practical.1 (SD = 1) and PANCE = 0.2907403.
## 
## 13. Correlation between MSK.CAT (SD = 1) and PANCE = 0.3251176.
## 
## 14. Correlation between PA731.Muscskl.Ds...Injury.Final.Grade (SD = 1) and PANCE = 0.3713423.
## 
## 15. Correlation between Neuro.iRAT.Mean (SD = 1) and PANCE = 0.4803686.
## 
## 16. Correlation between Neuro.Anatomy.Practical (SD = 1) and PANCE = 0.197391.
## 
## 17. Correlation between Neuro.CAT (SD = 1) and PANCE = 0.3788714.
## 
## 18. Correlation between PA734..Neurology.Final.Grade (SD = 1) and PANCE = 0.4156433.
## 
## 19. Correlation between Patient.Care.Final.Written.Exam..CAT.1. (SD = 1) and PANCE = 0.416525.
## 
## 20. Correlation between Patient.Care.I.Final.Grade (SD = 1) and PANCE = 0.2582716.
## 
## 21. Correlation between PA.in.Practice.Written.Exam.1 (SD = 1) and PANCE = 0.07682171.
## 
## 22. Correlation between PA.in.Practice..Practical.Exam.2 (SD = 1) and PANCE = 0.0001144805.
## 
## 23. Correlation between PA.in.Practice.Written.Exam.2 (SD = 1) and PANCE = 0.02831808.
```

```
## 
## 24. Correlation between PA721..PA.in.Practice..Final.Grade (SD = 1) and PANCE = 0.03645805.
## 
## 25. Correlation between Cardio.iRAT.Mean (SD = 1) and PANCE = 0.5121714.
## 
## 26. Correlation between Cardio.CAT..1 (SD = 1) and PANCE = 0.4998659.
## 
## 27. Correlation between PA736.Cardiovascular.Disease.Final.Grade (SD = 1) and PANCE = 0.5350263.
## 
## 28. Correlation between Pulm.iRAT.Mean (SD = 1) and PANCE = 0.4573783.
## 
## 29. Correlation between Pulm.CAT (SD = 1) and PANCE = 0.4965562.
## 
## 30. Correlation between PA733.Pulmonary.Med.Final.Grade (SD = 1) and PANCE = 0.5404144.
## 
## 31. Correlation between Otolar...Ophthal.iRAT.Mean (SD = 1) and PANCE = 0.4394858.
## 
## 32. Correlation between Otolar...OphthalAnatomy.Practical (SD = 1) and PANCE = 0.1584551.
## 
## 33. Correlation between Otolar...OphthalCAT (SD = 1) and PANCE = 0.2079298.
## 
## 34. Correlation between PA737.Otolar...Ophthal.Final.Grade (SD = 1) and PANCE = 0.3156686.
## 
## 35. Correlation between Behav.Med.iRAT.Mean (SD = 1) and PANCE = 0.3885394.
## 
## 36. Correlation between Behav.Med.CAT (SD = 1) and PANCE = 0.2893295.
## 
## 37. Correlation between PA735.Prin.of.Beh..Med.Final.Grade (SD = 1) and PANCE = 0.3518571.
## 
## 38. Correlation between Patient.Care.II.Exam.1 (SD = 1) and PANCE = 0.2616139.
## 
## 39. Correlation between PA751.Patient.Care.II.Final.Grade (SD = 1) and PANCE = 0.2672445.
## 
## 40. Correlation between Neph.Ur.iRAT.Mean (SD = 1) and PANCE = 0.2687676.
## 
## 41. Correlation between Neph.Uro.CAT (SD = 1) and PANCE = 0.2755651.
## 
## 42. Correlation between Endo.Anatomy.Practical (SD = 1) and PANCE = 0.1715459.
## 
## 43. Correlation between Endo.CAT (SD = 1) and PANCE = 0.3668406.
## 
## 44. Correlation between PA739..Endocrinology.Final.Grade (SD = 1) and PANCE = 0.4201746.
## 
## 45. Correlation between GI.iRAT.Mean (SD = 1) and PANCE = 0.2251731.
## 
## 46. Correlation between GI.CAT (SD = 1) and PANCE = 0.3629115.
## 
## 47. Correlation between GI.Anatomy.Practical (SD = 1) and PANCE = 0.08289179.
## 
## 48. Correlation between PA738.Gastroenterlogy.Final.Grade (SD = 1) and PANCE = 0.4150282.
## 
## 49. Correlation between Repro.iRAT.Mean (SD = 1) and PANCE = 0.431119.
## 
## 50. Correlation between Repro.Anatomy.Practical (SD = 1) and PANCE = 0.1448511.
```

```
## 
## 51. Correlation between Repro.CAT (SD = 1) and PANCE = 0.4604165.
## 
## 52. Correlation between PA741.Prin..of.Repro.Med.Final.Grade (SD = 1) and PANCE = 0.3940514.
## 
## 53. Correlation between Pt.Care.III.Exam.1 (SD = 1) and PANCE = 0.2108213.
## 
## 54. Correlation between PA752.Patient.Care.III.Final.Grade (SD = 1) and PANCE = 0.2629877.
## 
## 55. Correlation between Special.Pops.Midterm (SD = 1) and PANCE = 0.2966964.
## 
## 56. Correlation between Special.Pops.Final (SD = 1) and PANCE = 0.278615.
## 
## 57. Correlation between PA760.Special.Pops.Final.Grade (SD = 1) and PANCE = 0.3704975.
## 
## 58. Correlation between SEI.iRAT.Mean (SD = 1) and PANCE = 0.4279774.
## 
## 59. Correlation between SEI.CAT (SD = 1) and PANCE = 0.5089996.
## 
## 60. Correlation between PA770.Surg..EM..and.IP.Care.Final.Grade (SD = 1) and PANCE = 0.5675375.
## 
## 61. Correlation between PACKRAT.I..Raw.Score (SD = 1) and PANCE = 0.6443328.
## 
## 62. Correlation between PANCE (SD = 64.05958) and PANCE = 1.
## 
## Variable selection process is complete. Returned dataframe contains 181 observations and 52 variables
```

Make a copy of the new data set, with correlation-selected variables:

```r
FinalAlumni.b <- df
```

```r
names(FinalAlumni.b)
```

```
##  [1] "Overall.GPA"
##  [2] "Overall.Science.GPA"
##  [3] "Foundations.IRAT.Mean"
##  [4] "Foundations.CAT"
##  [5] "PA650.Foundations..of.Med.Final.Grade"
##  [6] "Derm.iRAT.Mean"
##  [7] "Derm.CAT"
##  [8] "PA730.Derm.Final.Grade"
##  [9] "MSK.iRAT.Mean"
## [10] "Practical.1"
## [11] "MSK.CAT"
## [12] "PA731.Muscskl.Ds...Injury.Final.Grade"
## [13] "Neuro.iRAT.Mean"
## [14] "Neuro.Anatomy.Practical"
## [15] "Neuro.CAT"
## [16] "PA734..Neurology.Final.Grade"
## [17] "Patient.Care.Final.Written.Exam..CAT.1."
## [18] "Patient.Care.I.Final.Grade"
## [19] "Cardio.iRAT.Mean"
## [20] "Cardio.CAT..1"
```

```
## [21] "PA736.Cardiovascular.Disease.Final.Grade"
## [22] "Pulm.iRAT.Mean"
## [23] "Pulm.CAT"
## [24] "PA733.Pulmonary.Med.Final.Grade"
## [25] "Otolar...Ophthal.iRAT.Mean"
## [26] "Otolar...OphthalCAT"
## [27] "PA737.Otolar...Ophthal.Final.Grade"
## [28] "Behav.Med.iRAT.Mean"
## [29] "Behav.Med.CAT"
## [30] "PA735.Prin.of.Beh..Med.Final.Grade"
## [31] "Patient.Care.II.Exam.1"
## [32] "PA751.Patient.Care.II.Final.Grade"
## [33] "Neph.Ur.iRAT.Mean"
## [34] "Neph.Uro.CAT"
## [35] "Endo.CAT"
## [36] "PA739..Endocrinology.Final.Grade"
## [37] "GI.iRAT.Mean"
## [38] "GI.CAT"
## [39] "PA738.Gastroenterlogy.Final.Grade"
## [40] "Repro.iRAT.Mean"
## [41] "Repro.CAT"
## [42] "PA741.Prin..of.Repro.Med.Final.Grade"
## [43] "Pt.Care.III.Exam.1"
## [44] "PA752.Patient.Care.III.Final.Grade"
## [45] "Special.Pops.Midterm"
## [46] "Special.Pops.Final"
## [47] "PA760.Special.Pops.Final.Grade"
## [48] "SEI.iRAT.Mean"
## [49] "SEI.CAT"
## [50] "PA770.Surg..EM..and.IP.Care.Final.Grade"
## [51] "PACKRAT.I..Raw.Score"
## [52] "PANCE"
```

```r
CorSelectedVars.b <- names(FinalAlumni.b)
saveRDS(CorSelectedVars.b, file = "CorSelectedVars.b.rds")
```

## 2.4 Prepare current student data

### 2.4.1 Details and initial setup

Import and prepare:

```r
current <- read_excel("AI Research Template downloaded 20220207.xlsx")
names(current) <- make.names(names(current))

current.copy <- current

library(jtools)
current<-jtools::standardize(current)

current$startYear <- current.copy$Year.student.began.program
current$PANCE <- current.copy$PANCE
```

```r
# alumni <- d.newpred[which(d.newpred$startYear<yearcutoff),]

# dim(alumni)
#
# alumni$startYear <- NULL
#
# alumni.x <- subset(alumni, select = -PANCE)
# dim(alumni.x)
# # alumni<-jtools::standardize(alumni)
#
# alumni.y <- alumni[c("PANCE")]
# dim(alumni.y)

alumniForCurrentPredictions <- current[which(current$startYear<yearcutoff),]
current <- current[which(current$startYear==yearcutoff),]
dim(current)
```

```
## [1]  46 427
```

```r
# current<- na.omit(current)
# dim(current)

current.copy <- current # current.copy gets reassigned here

# current$PANCE <- NULL
# current$startYear <- NULL
# current<-jtools::standardize(current)

# dim(current)
# names(current)
```

Now we have a dataframe called `current` which contains the latest graduating cohort only, meaning the *current students* on whom we need to make predictions. We also have other dataframes containing alumni data, which is all of the previous students on whom we will do LOOCV to build and compare models and with whom we will then train a model to run on the cohort of current students.

Below, two versions of the dataframe `current` will be made, each with the variables in either preparation A or B.

We also made a dataframe called `alumniForCurrentPredictions`. Two versions of this will also be madde below for preparations A and B. We cannot just use `FinalAlumni.a` and `FinalAlumni.b` for the validation on current students because those files were standardized *without* the inclusion of the 2019 current student validation cohort. **Standardization must occur when both the training and testing data are together**—*before* they are divided—which is what we achieve in this section, in preparation for running validation models later in this appendix.

```r
dim(current)
```

```
## [1]  46 427
```

```r
with(current, summary(PANCE))
```

```
##    Min. 1st Qu.  Median    Mean 3rd Qu.    Max.    NA's
##   305.0   376.5   433.0   434.0   476.0   615.0       3
```

Above, we see that there are 3 students in the current cohort who have not yet done the PANCE exam, meaning their dependent variable is missing.

We eliminate these people:

```
current <- current[which(!is.na(current$PANCE)),]
dim(current)
```

```
## [1]  43 427
```

```
with(current, summary(PANCE))
```

```
##    Min. 1st Qu.  Median    Mean 3rd Qu.    Max.
##   305.0   376.5   433.0   434.0   476.0   615.0
```

### 2.4.2 Preparation A – current students

```
CorSelectedVars.a <- readRDS("CorSelectedVars.a.rds")
current.a <- current[CorSelectedVars.a]
alumniForCurrentPredictions.a <- alumniForCurrentPredictions[CorSelectedVars.a]
```

```
dim(current.a)
```

```
## [1] 43 36
```

```
current.a <- na.omit(current.a)
dim(current.a)
```

```
## [1] 43 36
```

```
dim(alumniForCurrentPredictions.a)
```

```
## [1] 183  36
```

```
alumniForCurrentPredictions.a <- na.omit(alumniForCurrentPredictions.a)
dim(alumniForCurrentPredictions.a)
```

```
## [1] 181  36
```

```
current.a.finalcopy <- current.a
```

### 2.4.3 Preparation B – current students

```
CorSelectedVars.b <- readRDS("CorSelectedVars.b.rds")
current.b <- current[CorSelectedVars.b]
alumniForCurrentPredictions.b <- alumniForCurrentPredictions[CorSelectedVars.b]
```

```r
dim(current.b)
```

```
## [1] 43 52
```

```r
current.b <- na.omit(current.b)
dim(current.b)
```

```
## [1] 42 52
```

```r
dim(alumniForCurrentPredictions.b)
```

```
## [1] 183  52
```

```r
alumniForCurrentPredictions.b <- na.omit(alumniForCurrentPredictions.b)
dim(alumniForCurrentPredictions.b)
```

```
## [1] 181  52
```

```r
current.b.finalcopy <- current.b
```

# 3 Supplemental information on predictive modeling approach

Main details of our predictive modeling approach are in our main article text. This section contains some supplementary details.

## 3.1 Variable selection

The independent variables that we manually select and then give to the computer to further narrow down are shown in outputs within this document. Here, we provide a few additional notes regarding our independent variable preparation

- iRAT quiz scores: As noted in our article, we took the mean of all iRATs within each course for each student (separately for each course), and included that mean iRAT score variable for each course for eligibility in our final model. We did attempt to use individual iRAT scores as separate independent variables in other modeling attempts (not shown) and did not find these to be useful. Just like averaging a patient's blood pressure to get a true picture of their trends and tendencies, averaging iRATs seems to give a better view of a student's academic and test-taking abilities. Furthermore, since iRAT quizzes are given many times to the students each week, they stand to be changed from year to year (both in content and number). Since our goals rely upon data that can be compared year to year, using the mean of the iRAT scores within each course for each student is a simple way to make sure that there will not be any missing data but that we can still include daily or weekly iRAT data as a metric of formative assessment.

- In a few cases, there are courses in the curriculum that were not taken by all students in all years. We did not use data from such courses.

## 3.2  Key terminology

- AMMKNN – Adaptive minimum match k-nearest neighbors. This is the predictive approach that we developed ourselves—a modification to standard K-nearest neighbors—to provide more accurate and sensitive predictions for our data and applied context.

- KNN – K-nearest neighbors. A standard and traditional, matching-based approach to generating predictions.

- RF – Random Forest. A standard and traditional machine learning approach that is based on collections of decision trees.

- SVM – Support Vector Machine. A standard and traditional machine learning approach that draws boundaries between observations to make predictions.

- LOOCV – Leave-one-out cross validation. Defined in the main text of our article.

- PANCE – The dependent variable that we are trying to predict in our study. This is addressed in the main text of our article. Students must score 350 points or higher on PANCE to pass.

Running KNN, RF, and SVM are all standard practices within the field of educational predictive analytics as demonstrated by the literature we cite in the main text of our article.

LOOCV is a common evaluation approach for data with small sample sizes; we cite some of the literature that sets a precedent for this. We also go a step further than much of this literature by holding out our "current" 2019 cohort for further model validation.

## 3.3  Manual tuning and adjustment of model predictions (other than adaptive minimum match KNN)

In all of the predictive models that we created—other than the adaptive minimum match KNN procedure that we developed to combat this very problem—we find that the model over-estimates PANCE score (our dependent variable). This is highly visible in the results we present below in this appendix. When we take predictions from standard RF, SVM, and KNN models and compare them to students' actual results on the PANCE, the predictions are always too high. We present standard 2-by-2 confusion matrices for all of these scenarios.

When we take these standard models' predictions and classify students predicted to score below 350 as failing and 350 or above as passing, many models are unable to correctly predict even a single student who truly failed the PANCE. To make these standard models' results more meaningful, we manually adjust the predictions made by these models. For example, if a standard KNN model predicts that a student will score 385 points on the PANCE, we recode that score of 385 to be less than 350, such that it now predicts a failing result.

The modification of the prediction threshold that separates failing and passing for the predicted exam outcomes, as described above, makes each model's results more sensitive, meaning that it will classify more students as positives, meaning those who will fail the exam. Overall, this is what we want (to identify those who will fail), but the problem is that increasing the sensitivity also leads to more false positive predictions. In most cases, as shown below in the results, the number of false positives are too many for the model to be usable. Our best predictive approach, AMMKNN, also suffers from false positive predictions, but much fewer than the other approaches. AMMKNN achieves high sensitivity while still keeping the number of false positives within tolerable bounds, as the results later in this file show and as we discuss in the main article.

Our code and results show both untuned results and the modifications we made. As the results show, the results are only meaningful when we treat 390 and often even higher numbers as failing grades. Note that these adjustments are only being made to the predictions of student scores. The true values of their scores are never adjusted, of course. Also note that changing the cutoff between failing and passing from 350 to 390 for predictions is similar to simply subtracting 40 points from everybody's predicted scores.

Our code shows our attempts to fine-tune the results with different artificial cutoff thresholds for the predictions of standard machine learning models and how those yield different results. AMMKNN does not require this type of fine-tuning.

# 4 Supplemental information on evaluation of predictive models

Main details of our evaluation approach of our predictive models are in our main article. This section contains some supplementary details.

## 4.1 Evaluation metrics

Here is how we define all of our key evaluation metrics:

- True positives (TP)- The number of students who truly failed PANCE and were correctly predicted by the model to fail PANCE. To us as educators, these are "predictable support" students who truly need additional remedial support and our model succeeded in identifying them as such.

- False positives (FP)- The number of students who truly passed PANCE but were incorrectly predicted to fail. These are "unnecessary support" students who do not truly need our support but our model tells us that they do need remediation.

- True negatives (TN)- The number of students who truly passed PANCE and were correctly predicted to pass. These "predictable no support" students did not need our support and the model correctly predicted this.

- False negatives (FN)- The number of students who truly failed PANCE but were incorrectly predicted to pass. These students "fell through the cracks" because the model would ideally identify them as needing remedial support, but it did not. We argue that it is our duty to reduce this number as much as possible.

- Accuracy- The total number of correct predictions divided by the total number of students, (TP+TN)/181. Sensitivity- The proportion of students who truly failed who were correctly predicted to fail, TP/(TP+FN).

- Specificity- The proportion of students who truly passed who were correctly predicted to pass, TN/(TN+FP).

**Note** that we define "positive" and "negative" outcomes this way to maintain consistency with the use of these terms in healthcare: A positive outcome is unwanted, like failing a test or having a disease; a negative outcome is desired, meaning passing a test or not having a disease.

## 4.2 Use of LOOCV and "current" cohort validation

When developing/testing our predictive models, we attempted to follow the same approach that we would use when applying predictive modeling in practice. In practice, we plan to use the following procedure:

1. Use alumni data to train predictive models.
2. Determine if predictive models on alumni data are accurate, sensitive, and specific enough to trust to make predictions on new cohorts of students.

3. Use the best predictive model(s) to make predictions on new students.

Following this procedure, we treat our first four cohorts of students' data (2015–2018 starting cohorts) as alumni data and we treat our most recent cohort (starting in 2019) as the "current" cohort of students. We train and test (using LOOCV) the best model that we can on the alumni cohorts. We then use the best model to make predictions *all at once* on the *entire* 2019 "current" cohort, as if we had done so prior to their taking the PANCE. These two levels of testing/validation (first LOOCV among alumni and then full-cohort validation) show that the model has the potential to hold up and be useful in practice, rather than just as a model-building exercise.

All of the models we build are first tested with LOOCV and then validated on the "current" students cohort, with varying results that are shown below. AMMKNN performs the best for our purposes (best detection of failing and at-risk students in both LOOCV and full-cohort validation), with standard KNN coming close if it is heavily manually fine-tuned/adjusted.

In some cases, we are able to find a potentially-useful result from a standard (non-AMMKNN) model in our current cohort validation, but these often come from trial and error within the current cohort validation itself (instead of following directly from an LOOCV result), meaning that we would have no way of producing this result in practice in the future (when we do not yet know the true PANCE outcomes for the current cohort at the time when we generate the predictions and identify at-risk students).

# 5 Random forest

## 5.1 With Preparation A variables

### 5.1.1 LOOCV on alumni – Preparation A

```
library(randomForest)

df.rf.LOOCV <-
  crossValidate(FinalAlumni.a,
  "rf1 <- randomForest(PANCE ~ ., data=traindata, proximity=TRUE,ntree=1000);
  testdata$rf.pred <- predict(rf1, newdata = testdata)",
  nrow(FinalAlumni.a))
```

Inspect results:

```
df.rf.LOOCV$PassPANCE <- ifelse(df.rf.LOOCV$PANCE>349,1,0)

df.rf.LOOCV$rf1.pred.PassPANCE <-
  ifelse(df.rf.LOOCV$rf.pred > 349, 1, 0)

(cm1 <- with(df.rf.LOOCV,table(PassPANCE,rf1.pred.PassPANCE)))
```

```
##          rf1.pred.PassPANCE
## PassPANCE  0   1
##         0  0  13
##         1  1 167
```

Above, we see that when we take the results of the RF model as-is and do not make any modifications, it **does not correctly predict any of the 13 students** who failed the PANCE.

Manual fine-tuning/adjustment is needed to make the results even remotely useful. Below, we change the pass-fail threshold (only for the predicted values, of course) to 390 (rather than 349 like above):

```
df.rf.LOOCV$PassPANCE <- ifelse(df.rf.LOOCV$PANCE>349,1,0)

df.rf.LOOCV$rf1.pred.PassPANCE <- ifelse(df.rf.LOOCV$rf.pred > 390, 1, 0)

(cm1 <- with(df.rf.LOOCV,table(PassPANCE,rf1.pred.PassPANCE)))
```

```
##          rf1.pred.PassPANCE
## PassPANCE   0   1
##         0   5   8
##         1   6 162
```

Above, we now see that the model is more sensitive now and is able to detect 5 of the students who failed PANCE. But this is still a very low number, and this change also led to 8 false positive predictions. Therefore, this RF model is not practical to use or explore further. Below, we raise the prediction threshold even higher, which improves the predictions for those who truly failed, but also leads to many more false positives.

```
df.rf.LOOCV$rf1.pred.PassPANCE <- ifelse(df.rf.LOOCV$rf.pred > 400, 1, 0)

(cm1 <- with(df.rf.LOOCV,table(PassPANCE,rf1.pred.PassPANCE)))
```

```
##          rf1.pred.PassPANCE
## PassPANCE   0   1
##         0  10   3
##         1  25 143
```

```
df.rf.LOOCV$rf1.pred.PassPANCE <- ifelse(df.rf.LOOCV$rf.pred > 410, 1, 0)

(cm1 <- with(df.rf.LOOCV,table(PassPANCE,rf1.pred.PassPANCE)))
```

```
##          rf1.pred.PassPANCE
## PassPANCE   0   1
##         0  10   3
##         1  54 114
```

```
df.rf.LOOCV$rf1.pred.PassPANCE <- ifelse(df.rf.LOOCV$rf.pred > 420, 1, 0)

(cm1 <- with(df.rf.LOOCV,table(PassPANCE,rf1.pred.PassPANCE)))
```

```
##          rf1.pred.PassPANCE
## PassPANCE  0  1
##         0 11  2
##         1 76 92
```

### 5.1.2  Validate on current cohort – Preparation A

```r
current.a <- current.a.finalcopy

rf1 <- randomForest(PANCE ~ .,data=alumniForCurrentPredictions.a, proximity=TRUE,ntree=1000)
```

Inspect results:

```r
current.a$rf.pred.current.a <- predict(rf1, newdata = current.a)

current.a$PassPANCE <- ifelse(current.a$PANCE>349,1,0)

current.a$rf1.pred.current.a.PassPANCE <- ifelse(current.a$rf.pred.current.a > 349, 1, 0)

(cm1 <- with(current.a,table(PassPANCE,rf1.pred.current.a.PassPANCE)))
```

```
##          rf1.pred.current.a.PassPANCE
## PassPANCE  1
##         0  6
##         1 37
```

```r
current.a$rf1.pred.current.a.PassPANCE <- ifelse(current.a$rf.pred.current.a > 390, 1, 0)

(cm1 <- with(current.a,table(PassPANCE,rf1.pred.current.a.PassPANCE)))
```

```
##          rf1.pred.current.a.PassPANCE
## PassPANCE  0  1
##         0  0  6
##         1  1 36
```

```r
current.a$rf1.pred.current.a.PassPANCE <- ifelse(current.a$rf.pred.current.a > 400, 1, 0)

(cm1 <- with(current.a,table(PassPANCE,rf1.pred.current.a.PassPANCE)))
```

```
##          rf1.pred.current.a.PassPANCE
## PassPANCE  0  1
##         0  1  5
##         1  1 36
```

```r
current.a$rf1.pred.current.a.PassPANCE <- ifelse(current.a$rf.pred.current.a > 410, 1, 0)

(cm1 <- with(current.a,table(PassPANCE,rf1.pred.current.a.PassPANCE)))
```

```
##          rf1.pred.current.a.PassPANCE
## PassPANCE  0  1
##         0  1  5
##         1  2 35
```

```r
current.a$rf1.pred.current.a.PassPANCE <- ifelse(current.a$rf.pred.current.a > 420, 1, 0)

(cm1 <- with(current.a,table(PassPANCE,rf1.pred.current.a.PassPANCE)))
```

```
##           rf1.pred.current.a.PassPANCE
## PassPANCE  0  1
##         0  1  5
##         1  4 33
```

Above, we see that if we validate on our `current` students and manually make our model so sensitive that anyone predicted by the random forest model to score 420 has their score reduced to 350, then we can detect 1 of the 6 students who fail the exam. This comes at the expense of 4 false positives.

## 5.2 With Preparation B variables

### 5.2.1 LOOCV on alumni – Preparation B

```
library(randomForest)

df.rf.LOOCV <-
  crossValidate(FinalAlumni.b,
  "rf1 <- randomForest(PANCE ~ ., data=traindata, proximity=TRUE,ntree=1000);
  testdata$rf.pred <- predict(rf1, newdata = testdata)",
  nrow(FinalAlumni.b))
```

Inspect results:

```
df.rf.LOOCV$PassPANCE <- ifelse(df.rf.LOOCV$PANCE>349,1,0)

df.rf.LOOCV$rf1.pred.PassPANCE <-
  ifelse(df.rf.LOOCV$rf.pred > 349, 1, 0)

(cm1 <- with(df.rf.LOOCV,table(PassPANCE,rf1.pred.PassPANCE)))
```

```
##           rf1.pred.PassPANCE
## PassPANCE   1
##         0  13
##         1 168
```

```
df.rf.LOOCV$PassPANCE <- ifelse(df.rf.LOOCV$PANCE>349,1,0)

df.rf.LOOCV$rf1.pred.PassPANCE <- ifelse(df.rf.LOOCV$rf.pred > 390, 1, 0)

(cm1 <- with(df.rf.LOOCV,table(PassPANCE,rf1.pred.PassPANCE)))
```

```
##           rf1.pred.PassPANCE
## PassPANCE   0   1
##         0   5   8
##         1  11 157
```

```
df.rf.LOOCV$rf1.pred.PassPANCE <- ifelse(df.rf.LOOCV$rf.pred > 400, 1, 0)

(cm1 <- with(df.rf.LOOCV,table(PassPANCE,rf1.pred.PassPANCE)))
```

```
##           rf1.pred.PassPANCE
## PassPANCE   0   1
##          0 10   3
##          1 23 145
```

```r
df.rf.LOOCV$rf1.pred.PassPANCE <- ifelse(df.rf.LOOCV$rf.pred > 410, 1, 0)

(cm1 <- with(df.rf.LOOCV,table(PassPANCE,rf1.pred.PassPANCE)))
```

```
##           rf1.pred.PassPANCE
## PassPANCE   0   1
##          0 10   3
##          1 51 117
```

```r
df.rf.LOOCV$rf1.pred.PassPANCE <- ifelse(df.rf.LOOCV$rf.pred > 420, 1, 0)

(cm1 <- with(df.rf.LOOCV,table(PassPANCE,rf1.pred.PassPANCE)))
```

```
##           rf1.pred.PassPANCE
## PassPANCE   0   1
##          0 12   1
##          1 71  97
```

### 5.2.2 Validate on current cohort – Preparation B

```r
current.b <- current.b.finalcopy

rf1 <- randomForest(PANCE ~ .,data=alumniForCurrentPredictions.b, proximity=TRUE,ntree=1000)
```

Inspect results:

```r
current.b$rf.pred.current.b <- predict(rf1, newdata = current.b)

current.b$PassPANCE <- ifelse(current.b$PANCE>349,1,0)

current.b$rf1.pred.current.b.PassPANCE <- ifelse(current.b$rf.pred.current.b > 349, 1, 0)

(cm1 <- with(current.b,table(PassPANCE,rf1.pred.current.b.PassPANCE)))
```

```
##           rf1.pred.current.b.PassPANCE
## PassPANCE  1
##         0  6
##         1 36
```

```r
current.b$rf1.pred.current.b.PassPANCE <- ifelse(current.b$rf.pred.current.b > 390, 1, 0)

(cm1 <- with(current.b,table(PassPANCE,rf1.pred.current.b.PassPANCE)))
```

```
##          rf1.pred.current.b.PassPANCE
## PassPANCE  0  1
##         0  1  5
##         1  1 35
```

```r
current.b$rf1.pred.current.b.PassPANCE <- ifelse(current.b$rf.pred.current.b > 400, 1, 0)

(cm1 <- with(current.b,table(PassPANCE,rf1.pred.current.b.PassPANCE)))
```

```
##          rf1.pred.current.b.PassPANCE
## PassPANCE  0  1
##         0  1  5
##         1  1 35
```

```r
current.b$rf1.pred.current.b.PassPANCE <- ifelse(current.b$rf.pred.current.b > 410, 1, 0)

(cm1 <- with(current.b,table(PassPANCE,rf1.pred.current.b.PassPANCE)))
```

```
##          rf1.pred.current.b.PassPANCE
## PassPANCE  0  1
##         0  1  5
##         1  2 34
```

```r
current.b$rf1.pred.current.b.PassPANCE <- ifelse(current.b$rf.pred.current.b > 420, 1, 0)

(cm1 <- with(current.b,table(PassPANCE,rf1.pred.current.b.PassPANCE)))
```

```
##          rf1.pred.current.b.PassPANCE
## PassPANCE  0  1
##         0  2  4
##         1  4 32
```

Above, we see that if we validate on our `current` students and manually make our model so sensitive that anyone predicted by the random forest model to score 420 has their score reduced to 350, then we can detect 2 of the 6 students who fail the exam. This comes at the expense of 4 false positives.

# 6 SVM

## 6.1 With Preparation A variables

### 6.1.1 LOOCV on alumni – Preparation A

```r
library(e1071)

df.svm.LOOCV <-
  crossValidate(FinalAlumni.a,
                'svm1 <- svm(PANCE ~ ., data = traindata, kernel = "polynomial", gamma =1,
                cost = 10, scale = FALSE);
                testdata$svm1.pred <- predict(svm1, newdata = testdata)',
                nrow(FinalAlumni.a))
```

Inspect results:

```r
df.svm.LOOCV$PassPANCE <- ifelse(df.svm.LOOCV$PANCE>349,1,0)

df.svm.LOOCV$svm1.pred.PassPANCE <- ifelse(df.svm.LOOCV$svm1.pred >349, 1, 0)

(cm1 <- with(df.svm.LOOCV,table(PassPANCE,svm1.pred.PassPANCE)))
```

```
##          svm1.pred.PassPANCE
## PassPANCE   0   1
##         0   4   9
##         1   9 159
```

```r
df.svm.LOOCV$svm1.pred.PassPANCE <- ifelse(df.svm.LOOCV$svm1.pred > 390, 1, 0)

(cm1 <- with(df.svm.LOOCV,table(PassPANCE,svm1.pred.PassPANCE)))
```

```
##          svm1.pred.PassPANCE
## PassPANCE   0   1
##         0   8   5
##         1  18 150
```

```r
df.svm.LOOCV$svm1.pred.PassPANCE <- ifelse(df.svm.LOOCV$svm1.pred > 400, 1, 0)

(cm1 <- with(df.svm.LOOCV,table(PassPANCE,svm1.pred.PassPANCE)))
```

```
##          svm1.pred.PassPANCE
## PassPANCE   0   1
##         0  10   3
##         1  28 140
```

```r
df.svm.LOOCV$svm1.pred.PassPANCE <- ifelse(df.svm.LOOCV$svm1.pred > 410, 1, 0)

(cm1 <- with(df.svm.LOOCV,table(PassPANCE,svm1.pred.PassPANCE)))
```

```
##          svm1.pred.PassPANCE
## PassPANCE   0   1
##         0  10   3
##         1  37 131
```

```r
df.svm.LOOCV$svm1.pred.PassPANCE <- ifelse(df.svm.LOOCV$svm1.pred > 420, 1, 0)

(cm1 <- with(df.svm.LOOCV,table(PassPANCE,svm1.pred.PassPANCE)))
```

```
##          svm1.pred.PassPANCE
## PassPANCE   0   1
##         0  11   2
##         1  64 104
```

### 6.1.2 Validate on current cohort – Preparation A

```r
current.a <- current.a.finalcopy

svm1 <-
  svm(PANCE ~ ., data = alumniForCurrentPredictions.a, kernel = "polynomial", gamma=1, cost = 10, scale
```

Inspect results:

```r
current.a$svm1.pred.current.a <- predict(svm1, newdata = current.a)

current.a$PassPANCE <- ifelse(current.a$PANCE>349,1,0)

current.a$svm1.pred.current.a.PassPANCE <- ifelse(current.a$svm1.pred.current.a > 349, 1, 0)

(cm1 <- with(current.a,table(PassPANCE,svm1.pred.current.a.PassPANCE)))
```

```
##          svm1.pred.current.a.PassPANCE
## PassPANCE  0  1
##         0  1  5
##         1  2 35
```

```r
current.a$svm1.pred.current.a.PassPANCE <- ifelse(current.a$svm1.pred.current.a > 390, 1, 0)

(cm1 <- with(current.a,table(PassPANCE,svm1.pred.current.a.PassPANCE)))
```

```
##          svm1.pred.current.a.PassPANCE
## PassPANCE  0  1
##         0  2  4
##         1  3 34
```

```r
current.a$svm1.pred.current.a.PassPANCE <- ifelse(current.a$svm1.pred.current.a > 400, 1, 0)

(cm1 <- with(current.a,table(PassPANCE,svm1.pred.current.a.PassPANCE)))
```

```
##          svm1.pred.current.a.PassPANCE
## PassPANCE  0  1
##         0  2  4
##         1  4 33
```

```r
current.a$svm1.pred.current.a.PassPANCE <- ifelse(current.a$svm1.pred.current.a > 410, 1, 0)

(cm1 <- with(current.a,table(PassPANCE,svm1.pred.current.a.PassPANCE)))
```

```
##          svm1.pred.current.a.PassPANCE
## PassPANCE  0  1
##         0  3  3
##         1  5 32
```

```
current.a$svm1.pred.current.a.PassPANCE <- ifelse(current.a$svm1.pred.current.a > 420, 1, 0)

(cm1 <- with(current.a,table(PassPANCE,svm1.pred.current.a.PassPANCE)))
```

```
##          svm1.pred.current.a.PassPANCE
## PassPANCE  0  1
##         0  3  3
##         1  6 31
```

The SVM results above are not promising.

## 6.2 With Preparation B variables

### 6.2.1 LOOCV on alumni – Preparation B

```
library(e1071)

df.svm.LOOCV <-
  crossValidate(FinalAlumni.b,
                'svm1 <- svm(PANCE ~ ., data = traindata, kernel = "polynomial", gamma =1,
                cost = 10, scale = FALSE);
                testdata$svm1.pred <- predict(svm1, newdata = testdata)',
                nrow(FinalAlumni.b))
```

Inspect results:

```
df.svm.LOOCV$PassPANCE <- ifelse(df.svm.LOOCV$PANCE>349,1,0)

df.svm.LOOCV$svm1.pred.PassPANCE <- ifelse(df.svm.LOOCV$svm1.pred >349, 1, 0)

(cm1 <- with(df.svm.LOOCV,table(PassPANCE,svm1.pred.PassPANCE)))
```

```
##          svm1.pred.PassPANCE
## PassPANCE   0   1
##         0   4   9
##         1   7 161
```

```
df.svm.LOOCV$svm1.pred.PassPANCE <- ifelse(df.svm.LOOCV$svm1.pred > 390, 1, 0)

(cm1 <- with(df.svm.LOOCV,table(PassPANCE,svm1.pred.PassPANCE)))
```

```
##          svm1.pred.PassPANCE
## PassPANCE   0   1
##         0   9   4
##         1  18 150
```

Since the result above, with a 390 prediction threshold, has 9 true positives, we will look at it in a 3-by-3 matrix:

```r
`Actual Values` <- cut(df.svm.LOOCV$PANCE,
                       breaks=c(-Inf,350,375,Inf),
                       labels=c("<350","350-375",">375"))

`Predicted Values` <- cut(df.svm.LOOCV$svm1.pred,
                          breaks=c(-Inf,390,400,Inf),
                          labels=c("<350","350-375",">375"))

# (cm<- addmargins(table(`Actual Values`,`Predicted Values`)))

(cm1<- table(`Actual Values`,`Predicted Values`))
```

```
##              Predicted Values
## Actual Values <350 350-375 >375
##       <350      9       1    3
##       350-375   2       1   12
##       >375     16       5  132
```

```r
df.svm.LOOCV$svm1.pred.PassPANCE <- ifelse(df.svm.LOOCV$svm1.pred > 400, 1, 0)

(cm1 <- with(df.svm.LOOCV,table(PassPANCE,svm1.pred.PassPANCE)))
```

```
##          svm1.pred.PassPANCE
## PassPANCE   0   1
##         0  10   3
##         1  24 144
```

```r
df.svm.LOOCV$svm1.pred.PassPANCE <- ifelse(df.svm.LOOCV$svm1.pred > 410, 1, 0)

(cm1 <- with(df.svm.LOOCV,table(PassPANCE,svm1.pred.PassPANCE)))
```

```
##          svm1.pred.PassPANCE
## PassPANCE   0   1
##         0  10   3
##         1  36 132
```

```r
df.svm.LOOCV$svm1.pred.PassPANCE <- ifelse(df.svm.LOOCV$svm1.pred > 420, 1, 0)

(cm1 <- with(df.svm.LOOCV,table(PassPANCE,svm1.pred.PassPANCE)))
```

```
##          svm1.pred.PassPANCE
## PassPANCE   0   1
##         0  10   3
##         1  62 106
```

### 6.2.2 Validate on current cohort – Preparation B

```r
current.b <- current.b.finalcopy
```

```r
svm1 <-
  svm(PANCE ~ ., data = alumniForCurrentPredictions.b, kernel = "polynomial", gamma=1, cost = 10, scale
```

Inspect results:

```r
current.b$svm1.pred.current.b <- predict(svm1, newdata = current.b)

current.b$PassPANCE <- ifelse(current.b$PANCE>349,1,0)

current.b$svm1.pred.current.b.PassPANCE <- ifelse(current.b$svm1.pred.current.b > 349, 1, 0)

(cm1 <- with(current.b,table(PassPANCE,svm1.pred.current.b.PassPANCE)))
```

```
##          svm1.pred.current.b.PassPANCE
## PassPANCE  0  1
##         0  1  5
##         1  2 34
```

```r
current.b$svm1.pred.current.b.PassPANCE <- ifelse(current.b$svm1.pred.current.b > 390, 1, 0)

(cm1 <- with(current.b,table(PassPANCE,svm1.pred.current.b.PassPANCE)))
```

```
##          svm1.pred.current.b.PassPANCE
## PassPANCE  0  1
##         0  2  4
##         1  2 34
```

```r
current.b$svm1.pred.current.b.PassPANCE <- ifelse(current.b$svm1.pred.current.b > 400, 1, 0)

(cm1 <- with(current.b,table(PassPANCE,svm1.pred.current.b.PassPANCE)))
```

```
##          svm1.pred.current.b.PassPANCE
## PassPANCE  0  1
##         0  2  4
##         1  2 34
```

```r
current.b$svm1.pred.current.b.PassPANCE <- ifelse(current.b$svm1.pred.current.b > 410, 1, 0)

(cm1 <- with(current.b,table(PassPANCE,svm1.pred.current.b.PassPANCE)))
```

```
##          svm1.pred.current.b.PassPANCE
## PassPANCE  0  1
##         0  3  3
##         1  3 33
```

```r
current.b$svm1.pred.current.b.PassPANCE <- ifelse(current.b$svm1.pred.current.b > 420, 1, 0)

(cm1 <- with(current.b,table(PassPANCE,svm1.pred.current.b.PassPANCE)))
```

```
##         svm1.pred.current.b.PassPANCE
## PassPANCE  0  1
##         0  3  3
##         1  8 28
```

```r
`Predicted Values` <-
  cut(current.b$svm1.pred.current.b,
      breaks=c(-Inf,390,400,Inf),
      labels=c("<350","350-375",">375"))

`Actual Values` <-
  cut(current.b$PANCE,
      breaks=c(-Inf,350,375,Inf),
      labels=c("<350","350-375",">375"))

table(`Actual Values`,`Predicted Values`)
```

```
##              Predicted Values
## Actual Values <350 350-375 >375
##       <350       2       0    4
##       350-375    2       0    3
##       >375       0       0   31
```

These SVM results with Preparation B are better than with Preparation A, but we still have too low sensitivity (too few of those who truly fail are detected correctly) and too many false-positives.

# 7 KNN

Note that in a future section in this document, when we run adaptive minimum match KNN (AMMKNN), in the dataframe `adaKNNminMatch.LOOCV`, we have a variable called `EucMatchDVMean.12` which should be the predicted value for standard KNN with K=12 for each observation. And we also have a dataframe `adaKNNminMatch.current` with the variable `EucMatchDVMean.12`, which would be the predictions from KNN with K=12 for the validation cohort of students. So, making a KNN model using the `FNN` package is not essential, given that we will be making an AMMKNN model as well and our AMMKNN results include K=12 results for standard KNN, as explained.

However, for the sake of completeness, we present standard KNN in a more conventional way below.

## 7.1 With Preparation A variables

### 7.1.1 LOOCV on alumni – Preparation A

Train model with LOOCV, k = 12:

```r
library(dplyr)
library(FNN)

df.knn12.LOOCV <-
  crossValidate(FinalAlumni.a, 'dtrain.x <- traindata %>% select(-PANCE);
dtrain.y <- traindata %>% select(PANCE);
dtest.x <- testdata %>% select(-PANCE);
```

```r
dtest.y <- testdata %>% select(PANCE);
pred.a <- FNN::knn.reg(dtrain.x, dtest.x, dtrain.y, k = 12);
testdata$knn12.pred <- pred.a$pred', nrow(FinalAlumni.a))
```

Inspect results:

```r
df.knn12.LOOCV$PassPANCE <- ifelse(df.knn12.LOOCV$PANCE>349,1,0)

df.knn12.LOOCV$knn12.pred.PassPANCE <-
  ifelse(df.knn12.LOOCV$knn12.pred >349, 1, 0)

(cm1 <- with(df.knn12.LOOCV,table(PassPANCE,knn12.pred.PassPANCE)))
```

```
##          knn12.pred.PassPANCE
## PassPANCE   1
##         0  13
##         1 168
```

```r
df.knn12.LOOCV$knn12.pred.PassPANCE <-
  ifelse(df.knn12.LOOCV$knn12.pred > 390, 1, 0)

(cm1 <- with(df.knn12.LOOCV,table(PassPANCE,knn12.pred.PassPANCE)))
```

```
##          knn12.pred.PassPANCE
## PassPANCE   0   1
##         0   5   8
##         1   7 161
```

```r
df.knn12.LOOCV$knn12.pred.PassPANCE <-
  ifelse(df.knn12.LOOCV$knn12.pred > 400, 1, 0)

(cm1 <- with(df.knn12.LOOCV,table(PassPANCE,knn12.pred.PassPANCE)))
```

```
##          knn12.pred.PassPANCE
## PassPANCE   0   1
##         0   7   6
##         1  13 155
```

```r
df.knn12.LOOCV$knn12.pred.PassPANCE <-
  ifelse(df.knn12.LOOCV$knn12.pred > 410, 1, 0)

(cm1 <- with(df.knn12.LOOCV,table(PassPANCE,knn12.pred.PassPANCE)))
```

```
##          knn12.pred.PassPANCE
## PassPANCE   0   1
##         0   9   4
##         1  43 125
```

```r
df.knn12.LOOCV$knn12.pred.PassPANCE <-
  ifelse(df.knn12.LOOCV$knn12.pred > 420, 1, 0)

(cm1 <- with(df.knn12.LOOCV,table(PassPANCE,knn12.pred.PassPANCE)))
```

```
##          knn12.pred.PassPANCE
## PassPANCE   0    1
##         0  10    3
##         1  67  101
```

Since the result above—with the prediction cut-off at 400—is decent, we disaggregate:

```r
`Actual Values` <- cut(df.knn12.LOOCV$PANCE,
                       breaks=c(-Inf,350,375,Inf),
                       labels=c("<350","350-375",">375"))

`Predicted Values` <- cut(df.knn12.LOOCV$knn12.pred,
                          breaks=c(-Inf,400,405,Inf),
                          labels=c("<350","350-375",">375"))

# (cm<- addmargins(table(`Actual Values`,`Predicted Values`)))

(cm1<- table(`Actual Values`,`Predicted Values`))
```

```
##              Predicted Values
## Actual Values <350 350-375 >375
##       <350      7       1    5
##       350-375   2       3   10
##       >375     11      11  131
```

### 7.1.2 Validate on current cohort – Preparation A

Prepare training data:

```r
dtrain.x <- alumniForCurrentPredictions.a %>% select(-PANCE) # remove DV
dtrain.y <- alumniForCurrentPredictions.a %>% select(PANCE)  # keep only DV
dtrain.x <- as.data.frame(dtrain.x)
dtrain.y <- as.data.frame(dtrain.y)

# dtrain.x <- FinalAlumni.a %>% select(-PANCE) # remove DV
# dtrain.y <- FinalAlumni.a %>% select(PANCE) # keep only DV
```

Prepare validation data:

```r
current.a <- current.a.finalcopy

dtest.x <- current.a %>% select(-PANCE) # remove DV
dtest.y <- current.a %>% select(PANCE)  # keep only DV
```

Make predictions:

```r
library(FNN)
pred.a <- FNN::knn.reg(dtrain.x, dtest.x, dtrain.y, k = 12)
```

Inspect results:

```r
dtest.y$knn12.pred <- pred.a$pred

dtest.y$PassPANCE <- ifelse(dtest.y$PANCE>349,1,0)

dtest.y$knn12.pred.PassPANCE <- ifelse(dtest.y$knn12.pred > 349, 1, 0)

(cm1 <- with(dtest.y,table(PassPANCE,knn12.pred.PassPANCE)))
```

```
##          knn12.pred.PassPANCE
## PassPANCE  1
##         0  6
##         1 37
```

```r
dtest.y$knn12.pred.PassPANCE <- ifelse(dtest.y$knn12.pred > 390, 1, 0)

(cm1 <- with(dtest.y,table(PassPANCE,knn12.pred.PassPANCE)))
```

```
##          knn12.pred.PassPANCE
## PassPANCE  1
##         0  6
##         1 37
```

```r
dtest.y$knn12.pred.PassPANCE <- ifelse(dtest.y$knn12.pred > 400, 1, 0)

(cm1 <- with(dtest.y,table(PassPANCE,knn12.pred.PassPANCE)))
```

```
##          knn12.pred.PassPANCE
## PassPANCE  0  1
##         0  0  6
##         1  1 36
```

```r
dtest.y$knn12.pred.PassPANCE <- ifelse(dtest.y$knn12.pred > 410, 1, 0)

(cm1 <- with(dtest.y,table(PassPANCE,knn12.pred.PassPANCE)))
```

```
##          knn12.pred.PassPANCE
## PassPANCE  0  1
##         0  2  4
##         1  2 35
```

```r
dtest.y$knn12.pred.PassPANCE <- ifelse(dtest.y$knn12.pred > 420, 1, 0)

(cm1 <- with(dtest.y,table(PassPANCE,knn12.pred.PassPANCE)))
```

```
##           knn12.pred.PassPANCE
## PassPANCE  0  1
##         0  4  2
##         1  4 33
```

View results, 3x3:

```r
`Predicted Values` <-
  cut(dtest.y$knn12.pred,
      breaks=c(-Inf,415,420,Inf),
      labels=c("<350","350-375",">375"))

table(`Predicted Values`)
```

```
## Predicted Values
##    <350 350-375    >375
##       5       3      35
```

Make 3x3 confusion matrix, when actual values are known:

```r
`Actual Values` <-
  cut(dtest.y$PANCE,
      breaks=c(-Inf,350,375,Inf),
      labels=c("<350","350-375",">375"))

table(`Actual Values`,`Predicted Values`)
```

```
##              Predicted Values
## Actual Values <350 350-375 >375
##       <350       2       2    2
##       350-375    2       0    3
##       >375       1       1   30
```

The result above on the validation cohort of current students is actually quite good. But cutoff points as high as 410 or 420 with this model during LOOCV showed much too many false positive predictions, so there would be no way to know that cutoff points of 415 and 420 (the ones used above) might be useful for a cohort of new students (because in practice, true PANCE results will not be known for the current student cohort at the time when the model is applied; therefore the LOOCV results are the only guide we have regarding which model to use on a new cohort). Since there is no systematic way to arrive at the result above, it is not presented in our main article.

## 7.2 With Preparation B variables

### 7.2.1 LOOCV on alumni – Preparation B

Train model with LOOCV, k = 12:

```r
library(dplyr)
library(FNN)

df.knn12.LOOCV <-
```

```
crossValidate(FinalAlumni.b, 'dtrain.x <- traindata %>% select(-PANCE);
dtrain.y <- traindata %>% select(PANCE);
dtest.x <- testdata %>% select(-PANCE);
dtest.y <- testdata %>% select(PANCE);
pred.b <- FNN::knn.reg(dtrain.x, dtest.x, dtrain.y, k = 12);
testdata$knn12.pred <- pred.b$pred', nrow(FinalAlumni.b))
```

Inspect results:

```
df.knn12.LOOCV$PassPANCE <- ifelse(df.knn12.LOOCV$PANCE>349,1,0)

df.knn12.LOOCV$knn12.pred.PassPANCE <-
  ifelse(df.knn12.LOOCV$knn12.pred >349, 1, 0)

(cm1 <- with(df.knn12.LOOCV,table(PassPANCE,knn12.pred.PassPANCE)))
```

```
##          knn12.pred.PassPANCE
## PassPANCE   1
##         0  13
##         1 168
```

```
df.knn12.LOOCV$knn12.pred.PassPANCE <-
  ifelse(df.knn12.LOOCV$knn12.pred > 390, 1, 0)

(cm1 <- with(df.knn12.LOOCV,table(PassPANCE,knn12.pred.PassPANCE)))
```

```
##          knn12.pred.PassPANCE
## PassPANCE   0   1
##         0   6   7
##         1   3 165
```

```
df.knn12.LOOCV$knn12.pred.PassPANCE <-
  ifelse(df.knn12.LOOCV$knn12.pred > 400, 1, 0)

(cm1 <- with(df.knn12.LOOCV,table(PassPANCE,knn12.pred.PassPANCE)))
```

```
##          knn12.pred.PassPANCE
## PassPANCE   0   1
##         0   8   5
##         1  17 151
```

```
df.knn12.LOOCV$knn12.pred.PassPANCE <-
  ifelse(df.knn12.LOOCV$knn12.pred > 410, 1, 0)

(cm1 <- with(df.knn12.LOOCV,table(PassPANCE,knn12.pred.PassPANCE)))
```

```
##          knn12.pred.PassPANCE
## PassPANCE   0   1
##         0  10   3
##         1  39 129
```

```
df.knn12.LOOCV$knn12.pred.PassPANCE <-
  ifelse(df.knn12.LOOCV$knn12.pred > 420, 1, 0)

(cm1 <- with(df.knn12.LOOCV,table(PassPANCE,knn12.pred.PassPANCE)))
```

```
##          knn12.pred.PassPANCE
## PassPANCE  0   1
##         0 11   2
##         1 64 104
```

Since the result above is decent, we disaggregate:

```
`Actual Values` <- cut(df.knn12.LOOCV$PANCE,
                    breaks=c(-Inf,350,375,Inf),
                    labels=c("<350","350-375",">375"))

`Predicted Values` <- cut(df.knn12.LOOCV$knn12.pred,
                       breaks=c(-Inf,400,410,Inf),
                       labels=c("<350","350-375",">375"))

# (cm<- addmargins(table(`Actual Values`,`Predicted Values`)))

(cm1<- table(`Actual Values`,`Predicted Values`))
```

```
##              Predicted Values
## Actual Values <350 350-375 >375
##       <350       8       2    3
##       350-375    1       2   12
##       >375      16      20  117
```

Above—when k=12 and we manually change the prediction cutoffs to 400 and 410 for the predicted values—the predictions are not bad, but 16 students who truly score above 375 are predicted to fail, which is too many. Adaptive Minimum Match KNN—our selected best model—predicts fewer students in this category. Furthermore, the standard KNN result above required careful manual adjustment of the prediction cutoffs before the results were useful. Adaptive Minimum Match KNN doesn't require such adjustment. These characteristics were also present in the disaggregated confusion matrix when k was set to a variety of numbers, as low as 6 and as high as 18, for other standard KNN models (which are not shown).

The reason for the manually adjusted prediction cutoffs above is that the standard KNN model gives everybody or nearly everybody a predicted score above 350. By changing the cutoff for passing or failing from 350 to 400 or more, *we essentially subtract 50 points from each student's predicted score.* Only after this subtraction do the results have some utility (although there are still too many false-positive predictions).

Below, we try to disaggregate again, with cutoffs at 415 and 420:

```
`Actual Values` <- cut(df.knn12.LOOCV$PANCE,
                    breaks=c(-Inf,350,375,Inf),
                    labels=c("<350","350-375",">375"))

`Predicted Values` <- cut(df.knn12.LOOCV$knn12.pred,
                       breaks=c(-Inf,415,420,Inf),
                       labels=c("<350","350-375",">375"))
```

```r
# (cm<- addmargins(table(`Actual Values`,`Predicted Values`)))

(cm1<- table(`Actual Values`,`Predicted Values`))
```

```
##                Predicted Values
## Actual Values <350 350-375 >375
##       <350      11       0    2
##       350-375    7       4    4
##       >375      45       8  100
```

Above, we see that there are 45 false positives which is too many for this version of our standard KNN result to be viable.

### 7.2.2 Validate on current cohort – Preparation B

Prepare training data:

```r
dtrain.x <- alumniForCurrentPredictions.b %>% select(-PANCE) # remove DV
dtrain.y <- alumniForCurrentPredictions.b %>% select(PANCE) # keep only DV

dtrain.x <- as.data.frame(dtrain.x)
dtrain.y <- as.data.frame(dtrain.y)
```

Prepare validation data:

```r
current.b <- current.b.finalcopy

dtest.x <- current.b %>% select(-PANCE) # remove DV
dtest.y <- current.b %>% select(PANCE) # keep only DV
```

Make predictions:

```r
library(FNN)
pred.b <- FNN::knn.reg(dtrain.x, dtest.x, dtrain.y, k = 12)
```

Inspect results:

```r
dtest.y$knn12.pred <- pred.b$pred

dtest.y$PassPANCE <- ifelse(dtest.y$PANCE>349,1,0)

dtest.y$knn12.pred.PassPANCE <- ifelse(dtest.y$knn12.pred > 349, 1, 0)

(cm1 <- with(dtest.y,table(PassPANCE,knn12.pred.PassPANCE)))
```

```
##          knn12.pred.PassPANCE
## PassPANCE  1
##         0  6
##         1 36
```

```r
dtest.y$knn12.pred.PassPANCE <- ifelse(dtest.y$knn12.pred > 390, 1, 0)

(cm1 <- with(dtest.y,table(PassPANCE,knn12.pred.PassPANCE)))
```

```
##          knn12.pred.PassPANCE
## PassPANCE  0  1
##         0  0  6
##         1  1 35
```

```r
dtest.y$knn12.pred.PassPANCE <- ifelse(dtest.y$knn12.pred > 400, 1, 0)

(cm1 <- with(dtest.y,table(PassPANCE,knn12.pred.PassPANCE)))
```

```
##          knn12.pred.PassPANCE
## PassPANCE  0  1
##         0  2  4
##         1  1 35
```

```r
dtest.y$knn12.pred.PassPANCE <- ifelse(dtest.y$knn12.pred > 410, 1, 0)

(cm1 <- with(dtest.y,table(PassPANCE,knn12.pred.PassPANCE)))
```

```
##          knn12.pred.PassPANCE
## PassPANCE  0  1
##         0  3  3
##         1  4 32
```

```r
dtest.y$knn12.pred.PassPANCE <- ifelse(dtest.y$knn12.pred > 420, 1, 0)

(cm1 <- with(dtest.y,table(PassPANCE,knn12.pred.PassPANCE)))
```

```
##          knn12.pred.PassPANCE
## PassPANCE  0  1
##         0  4  2
##         1  5 31
```

View results, 3x3:

```r
`Predicted Values` <-
  cut(dtest.y$knn12.pred,
      breaks=c(-Inf,400,410,Inf),
      labels=c("<350","350-375",">375"))

table(`Predicted Values`)
```

```
## Predicted Values
##    <350 350-375    >375
##       3       4      35
```

Make 3x3 confusion matrix, when actual values are known:

```r
`Actual Values` <-
  cut(dtest.y$PANCE,
      breaks=c(-Inf,350,375,Inf),
      labels=c("<350","350-375",">375"))

table(`Actual Values`,`Predicted Values`)
```

```
##              Predicted Values
## Actual Values <350 350-375 >375
##       <350      2       1    3
##       350-375   1       1    3
##       >375      0       2   29
```

Now we repeat the table above with cutoffs for predictions at 415 and 420:

```r
`Predicted Values` <-
  cut(dtest.y$knn12.pred,
      breaks=c(-Inf,415,420,Inf),
      labels=c("<350","350-375",">375"))

`Actual Values` <-
  cut(dtest.y$PANCE,
      breaks=c(-Inf,350,375,Inf),
      labels=c("<350","350-375",">375"))

table(`Actual Values`,`Predicted Values`)
```

```
##              Predicted Values
## Actual Values <350 350-375 >375
##       <350      3       1    2
##       350-375   2       1    2
##       >375      2       0   29
```

The result above is decent when we change the cutoff values to 415 and 420, but there is no way that we would have known to do this before knowing the true PANCE scores of the current students.

# 8 Adaptive minimum match KNN

## 8.1 With Preparation A variables

### 8.1.1 LOOCV on alumni – Preparation A

Train model with LOOCV:

```r
library(dplyr)

adaKNNminMatch.LOOCV <-
  crossValidate(
    FinalAlumni.a,
    'dtrain.x <- traindata %>% select(-PANCE);
```

```r
    dtrain.y <- traindata %>% select(PANCE);
    dtest.x <- testdata %>% select(-PANCE);
    pred.a <- adaptiveMinMatchKNNregression(
    dtrain.x, dtrain.y, dtest.x, maxK = 20);
    testdata <- pred.a', nrow(FinalAlumni.a))
```

## Warning: package 'factoextra' was built under R version 4.0.5

## Warning: package 'ggplot2' was built under R version 4.0.5

- In this and many other situations, the value of maxK does not appear matter too much, as long as it is high enough. We suspect that this is because the mean of all matches will stabilize as the number of matches increases. In other words: the difference between the mean of the first 20 matches and the mean of the first 19 matches is likely smaller than the difference between the mean of the first 11 matches and the mean of the first 10 matches.

Inspect results:

```r
# Add DV into predictive results
adaKNNminMatch.LOOCV$PANCE <- FinalAlumni.a$PANCE

# Make true pass/fail DV column
adaKNNminMatch.LOOCV$PassPANCE <- ifelse(adaKNNminMatch.LOOCV$PANCE>349, 1,0)

# Make predicted DV column
# those who scored less than 2 SDs below the mean on PACKRAT I
# are predicted to score the minimum of all matches (with no
# averaging involved);
# Everyone else is predicted to score the minimum of means;
adaKNNminMatch.LOOCV$pred.adj.PANCE <-
  ifelse(
    adaKNNminMatch.LOOCV$PACKRAT.I..Raw.Score < -2,
    adaKNNminMatch.LOOCV$MatchDVMin,
    adaKNNminMatch.LOOCV$MatchDVMeanMin)

# Make predicted pass/fail DV column
adaKNNminMatch.LOOCV$pred.adj.PassPANCE <-
  ifelse(adaKNNminMatch.LOOCV$pred.adj.PANCE > 349, 1, 0)

(cm.2by2 <- with(adaKNNminMatch.LOOCV,table(PassPANCE,pred.adj.PassPANCE)))
```

```
##          pred.adj.PassPANCE
## PassPANCE   0   1
##         0   8   5
##         1  11 157
```

Accuracy:

```r
(cm.2by2[1,1]+cm.2by2[2,2])/(cm.2by2[1,1]+cm.2by2[1,2]+cm.2by2[2,1]+cm.2by2[2,2])
```

```
## [1] 0.9116022
```

Sensivitivy:

```r
cm.2by2[1,1]/(cm.2by2[1,1]+cm.2by2[1,2])
```

```
## [1] 0.6153846
```

Specificity:

```r
cm.2by2[2,2]/(cm.2by2[2,1]+cm.2by2[2,2])
```

```
## [1] 0.9345238
```

Change the prediction threshold (optional):

```r
adaKNNminMatch.LOOCV$pred.adj.PassPANCE <-
  ifelse(adaKNNminMatch.LOOCV$pred.adj.PANCE > 365, 1, 0)

(cm1.370 <- with(adaKNNminMatch.LOOCV,table(PassPANCE,pred.adj.PassPANCE))) # changed name of cm1 to cm
```

```
##          pred.adj.PassPANCE
## PassPANCE   0   1
##         0   9   4
##         1  20 148
```

Disaggregated prediction:

```r
`Actual Values` <-
  cut(
    adaKNNminMatch.LOOCV$PANCE,
    breaks=c(-Inf,350,375,Inf),
    labels=c("<350","350-375",">375"))

`Predicted Values` <-
  cut(adaKNNminMatch.LOOCV$pred.adj.PANCE,
      breaks=c(-Inf,350,375,Inf),
      labels=c("<350","350-375",">375"))

# (cm<- addmargins(table(`Actual Values`,`Predicted Values`)))

(cm1<- table(`Actual Values`,`Predicted Values`))
```

```
##                Predicted Values
## Actual Values <350 350-375 >375
##       <350       8       3    2
##       350-375    3       4    8
##       >375       8      17  128
```

### 8.1.2 Validate on current cohort – Preparation A

Prepare training data:

```
dtrain.x <- alumniForCurrentPredictions.a %>% select(-PANCE) # remove DV
dtrain.y <- alumniForCurrentPredictions.a %>% select(PANCE) # keep only DV
```

Prepare validation data:

```
current.a <- current.a.finalcopy

dtest.x <- current.a %>% select(-PANCE) # remove DV
dtest.y <- current.a %>% select(PANCE) # keep only DV
```

Make predictions:

```
adaKNNminMatch.current <-
  adaptiveMinMatchKNNregression(dtrain.x, dtrain.y, dtest.x, maxK = 20)
```

Process predictions (penalize for low PACKRAT scores):

```
adaKNNminMatch.current$pred.adj.PANCE <-
  ifelse(
    adaKNNminMatch.current$PACKRAT.I..Raw.Score < -2,
    adaKNNminMatch.current$MatchDVMin,
    adaKNNminMatch.current$MatchDVMeanMin)

# Make predicted pass/fail DV column
adaKNNminMatch.current$pred.adj.PassPANCE <-
  ifelse(adaKNNminMatch.current$pred.adj.PANCE > 349, 1, 0)
```

Add back true DV for evaluation:

```
adaKNNminMatch.current$True.PANCE <- dtest.y$PANCE
```

View results:

```
`Predicted Values` <-
  cut(adaKNNminMatch.current$pred.adj.PANCE,
      breaks=c(-Inf,350,385,Inf),
      labels=c("<350","350-375",">375"))

table(`Predicted Values`, useNA = "always")

## Predicted Values
##    <350 350-375    >375    <NA>
##       4       2      37       0
```

Make confusion matrix, if actual values are known:

```r
`Actual Values` <-
  cut(adaKNNminMatch.current$True.PANCE,
      breaks=c(-Inf,350,375,Inf),
      labels=c("<350","350-375",">375"))

table(`Actual Values`,`Predicted Values`, useNA = "always")
```

```
##              Predicted Values
## Actual Values <350 350-375 >375 <NA>
##       <350       2       2    2    0
##       350-375    0       0    5    0
##       >375       2       0   30    0
##       <NA>       0       0    0    0
```

Visualize results:

```r
cor(adaKNNminMatch.current$pred.adj.PANCE, adaKNNminMatch.current$True.PANCE)
```

```
## [1] 0.7589687
```

```r
plot(adaKNNminMatch.current$pred.adj.PANCE, adaKNNminMatch.current$True.PANCE, xlab = "Predicted PANCE"
abline(v=350, col="red")
abline(h=350, col="red")
abline(v=375, col="green")
abline(h=375, col="green")
```

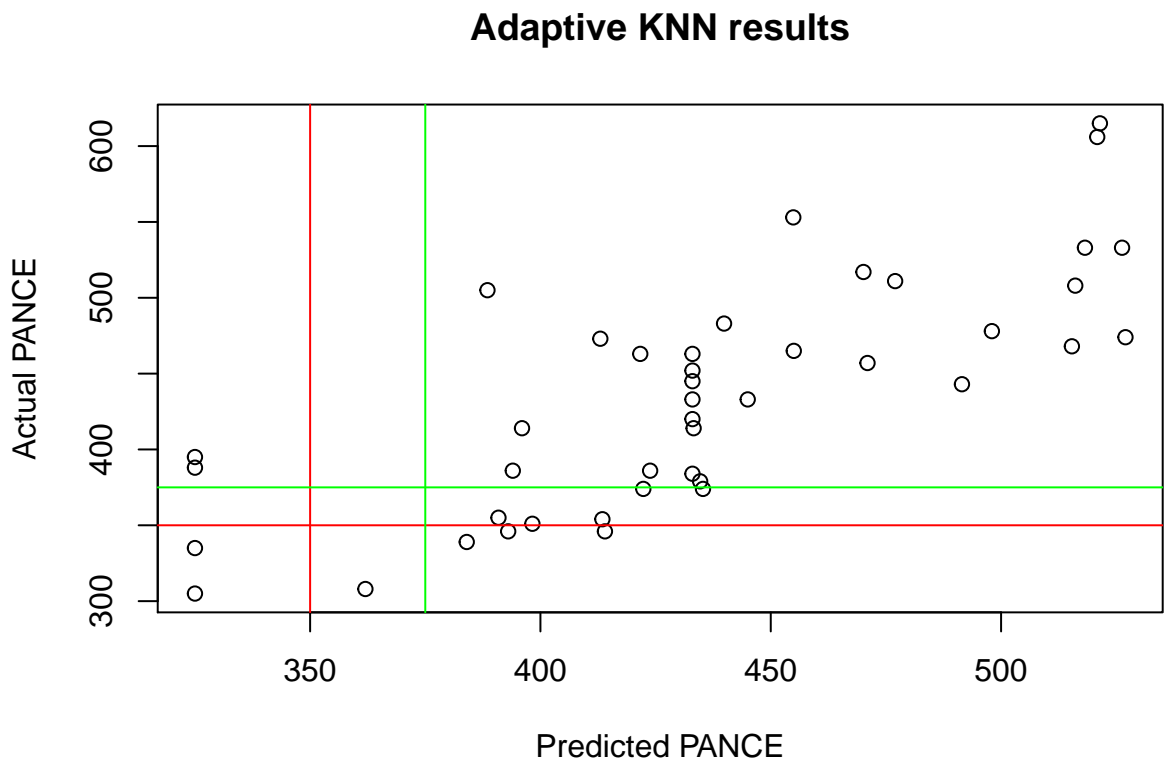

```
plot(adaKNNminMatch.current$PACKRAT.I..Raw.Score, adaKNNminMatch.current$True.PANCE, xlab = "PACKRAT I
abline(h=350, col="red")
abline(v=-0.5, col="red")
```

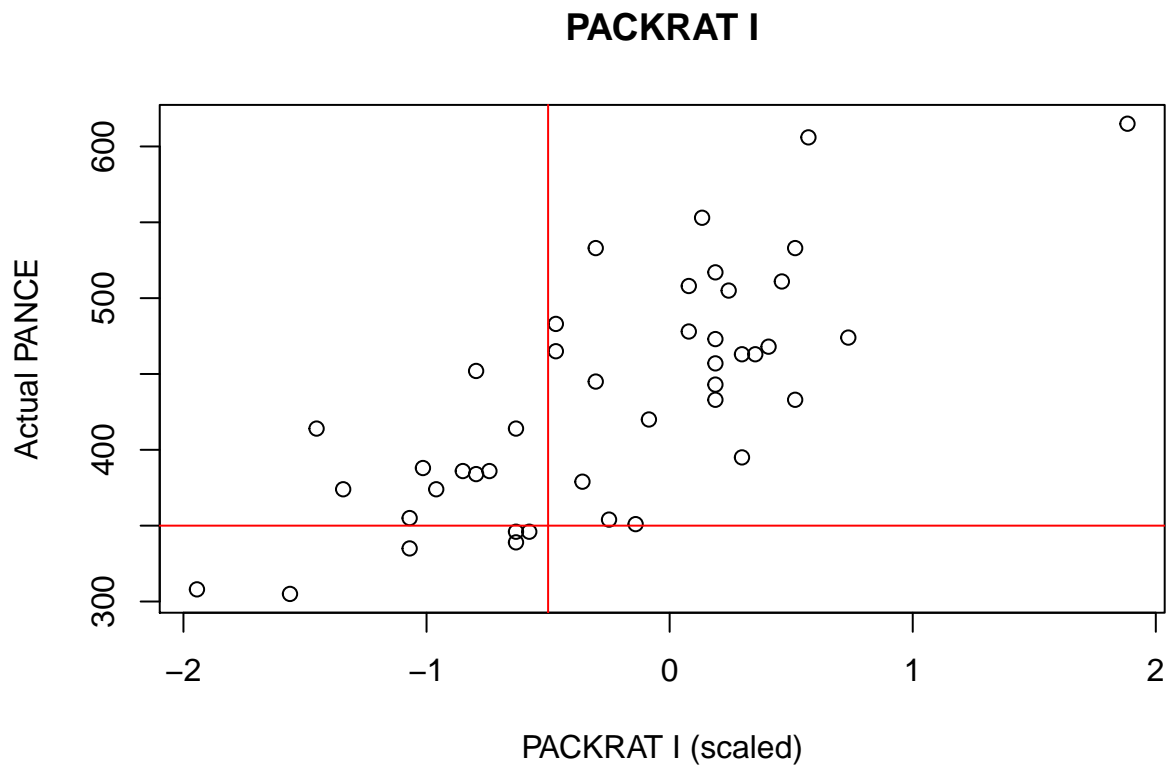

```
# abline(v=375, col="green")
# abline(h=375, col="green")
```

In the two scatterplots above, we see that adaptive minimum match KNN gives us more useful predictions than the PACKRAT (a nationally benchmarked cumulative exam commonly used to prepare for the PANCE). True PANCE results are shown on the vertical axis for the 2019 current student cohort. Predicted results (either from adaptive minimum match KNN or PACKRAT) are on the horizontal axis. Our goal is to draw a vertical line which has students who fail to the left and those who pass to the right. With adaptive minimum match KNN, we are able to draw this vertical line where the predicted PANCE score is equal to 375. This allows us to identify (to the left of the line) most of the students who fail without making too many mistakes. However, if we attempt to do the same on the second scatterplot with PACKRAT scores, it is not possible to draw a vertical line anywhere that identifies most of the students who fail without making too many wrong predictions.

## 8.2 With Preparation B variables

### 8.2.1 LOOCV on alumni – Preparation B

Train model with LOOCV:

```r
library(dplyr)

adaKNNminMatch.LOOCV <-
  crossValidate(
    FinalAlumni.b,
    'dtrain.x <- traindata %>% select(-PANCE);
    dtrain.y <- traindata %>% select(PANCE);
    dtest.x <- testdata %>% select(-PANCE);
    pred.b <- adaptiveMinMatchKNNregression(
    dtrain.x, dtrain.y, dtest.x, maxK = 20);
    testdata <- pred.b', nrow(FinalAlumni.b))
```

- In this and many other situations, the value of `maxK` does not appear matter too much, as long as it is high enough. We suspect that this is because the mean of all matches will stabilize as the number of matches increases. In other words: the difference between the mean of the first 20 matches and the mean of the first 19 matches is likely smaller than the difference between the mean of the first 11 matches and the mean of the first 10 matches.

Inspect results:

```r
# Add DV into predictive results
adaKNNminMatch.LOOCV$PANCE <- FinalAlumni.b$PANCE

# Make true pass/fail DV column
adaKNNminMatch.LOOCV$PassPANCE <- ifelse(adaKNNminMatch.LOOCV$PANCE>349, 1,0)

# Make predicted DV column
# those who scored less than 2 SDs below the mean on PACKRAT I
# are predicted to score the minimum of all matches (with no
# averaging involved);
# Everyone else is predicted to score the minimum of means;
adaKNNminMatch.LOOCV$pred.adj.PANCE <-
  ifelse(
    adaKNNminMatch.LOOCV$PACKRAT.I..Raw.Score < -2,
    adaKNNminMatch.LOOCV$MatchDVMin,
    adaKNNminMatch.LOOCV$MatchDVMeanMin)

# Make predicted pass/fail DV column
adaKNNminMatch.LOOCV$pred.adj.PassPANCE <-
  ifelse(adaKNNminMatch.LOOCV$pred.adj.PANCE > 349, 1, 0)

(cm.2by2 <- with(adaKNNminMatch.LOOCV,table(PassPANCE,pred.adj.PassPANCE)))
```

```
##          pred.adj.PassPANCE
## PassPANCE   0   1
##         0   9   4
##         1   9 159
```

Accuracy:

```r
(cm.2by2[1,1]+cm.2by2[2,2])/(cm.2by2[1,1]+cm.2by2[1,2]+cm.2by2[2,1]+cm.2by2[2,2])
```

```
## [1] 0.9281768
```

Sensivitivy:

```r
cm.2by2[1,1]/(cm.2by2[1,1]+cm.2by2[1,2])
```

```
## [1] 0.6923077
```

Specificity:

```r
cm.2by2[2,2]/(cm.2by2[2,1]+cm.2by2[2,2])
```

```
## [1] 0.9464286
```

Change the prediction threshold (optional):

```r
adaKNNminMatch.LOOCV$pred.adj.PassPANCE <-
  ifelse(adaKNNminMatch.LOOCV$pred.adj.PANCE > 365, 1, 0)

(cm1.370 <- with(adaKNNminMatch.LOOCV,table(PassPANCE,pred.adj.PassPANCE))) # changed name of cm1 to cm
```

```
##          pred.adj.PassPANCE
## PassPANCE   0   1
##         0  10   3
##         1  20 148
```

Disaggregated prediction:

```r
`Actual Values` <-
  cut(
    adaKNNminMatch.LOOCV$PANCE,
    breaks=c(-Inf,350,375,Inf),
    labels=c("<350","350-375",">375"))

`Predicted Values` <-
  cut(adaKNNminMatch.LOOCV$pred.adj.PANCE,
      breaks=c(-Inf,350,375,Inf),
      labels=c("<350","350-375",">375"))

# (cm<- addmargins(table(`Actual Values`,`Predicted Values`)))

(cm1<- table(`Actual Values`,`Predicted Values`))
```

```
##              Predicted Values
## Actual Values <350 350-375 >375
##       <350       9       2    2
##       350-375    1       3   11
##       >375       8      21  124
```

### 8.2.2 Validate on current cohort – Preparation B

Prepare training data:

```
dtrain.x <- alumniForCurrentPredictions.b %>% select(-PANCE) # remove DV
dtrain.y <- alumniForCurrentPredictions.b %>% select(PANCE) # keep only DV
```

Prepare validation data:

```
current.b <- current.b.finalcopy

dtest.x <- current.b %>% select(-PANCE) # remove DV
dtest.y <- current.b %>% select(PANCE) # keep only DV
```

Make predictions:

```
adaKNNminMatch.current <-
  adaptiveMinMatchKNNregression(dtrain.x, dtrain.y, dtest.x, maxK = 20)
```

Process predictions (penalize for low PACKRAT scores):

```
adaKNNminMatch.current$pred.adj.PANCE <-
  ifelse(
    adaKNNminMatch.current$PACKRAT.I..Raw.Score < -2,
    adaKNNminMatch.current$MatchDVMin,
    adaKNNminMatch.current$MatchDVMeanMin)

# Make predicted pass/fail DV column
adaKNNminMatch.current$pred.adj.PassPANCE <-
  ifelse(adaKNNminMatch.current$pred.adj.PANCE > 349, 1, 0)
```

Add back true DV for evaluation:

```
adaKNNminMatch.current$True.PANCE <- dtest.y$PANCE
```

View results:

```
`Predicted Values` <-
  cut(adaKNNminMatch.current$pred.adj.PANCE,
      breaks=c(-Inf,350,385,Inf),
      labels=c("<350","350-375",">375"))

table(`Predicted Values`, useNA = "always")

## Predicted Values
##    <350 350-375    >375   <NA>
##       5       6      31      0
```

Make confusion matrix, if actual values are known:

```r
`Actual Values` <-
  cut(adaKNNminMatch.current$True.PANCE,
      breaks=c(-Inf,350,375,Inf),
      labels=c("<350","350-375",">375"))

table(`Actual Values`,`Predicted Values`, useNA = "always")
```

```
##              Predicted Values
## Actual Values <350 350-375 >375 <NA>
##       <350       2       2    2    0
##       350-375    0       2    3    0
##       >375       3       2   26    0
##       <NA>       0       0    0    0
```

# 9 Study limitations and strengths

While our analytics approach is useful as an in-house tool and might be applicable to broader educational settings, we acknowledge a number of limitations in our study and results. The ability of our AMMKNN model to generalize to other types of health professions programs is unknown. Therefore, we recommend that others using this approach conduct multiple stages of validation after training the model and work to transparently apply the results when used for making predictions. Given our small sample size of five cohorts of students in a single program, we also cannot yet comment on whether our model's performance will improve or not after additional data are added.

The predictive model depends on collecting the same independent variables on students every year. If curriculum changes or instructional practices lead to non-comparable data being collected across cohorts, our predictive approach might not work as effectively. Furthermore, the cohorts included in our study involved some students that were and others who were not affected by the COVID-19 pandemic. Some students completed the PANCE exam in the middle of the pandemic, while most completed it prior to the pandemic. We do not currently measure and include variables to account for these potential cohort-specific differences. When we included a variable that identified specific cohorts, predictions did not improve, suggesting that some variation across cohorts might be accommodated by our approach.

Our study also has some strengths to build upon in future work. Educational analytics are often plagued by small sample sizes. Our AMMKNN method addresses this problem and was able to make improvements over standard models, without the need for extensive fine tuning of model parameters. Our ability to make reasonable predictions on an entire new cohort of students, to demonstrate application of the method, is another strength. The method now needs to be tested with other data and contexts. Many educational studies have cohorts in these size ranges and our methods reflected this common constraint.